\documentclass[preprint2,pdftex]{aastex}

\usepackage{amssymb}  
\usepackage{amsmath}  
\usepackage{mathrsfs} 
\usepackage{paralist} 

\newif\ifpdf
\ifx\pdfoutput\undefined
  \pdffalse
\else
  \ifnum\pdfoutput=1
    \pdftrue
  \else
    \pdffalse
  \fi
\fi

\pdftrue
\ifpdf
  \DeclareGraphicsExtensions{.pdf}
  \def\figdir{FIG.PDF}
\else
  \DeclareGraphicsExtensions{.eps}
  \def\figdir{./}
\fi

\newif{\ifrgb}
\rgbtrue

\newcommand{\TNSN}{TN\,SN}
\newcommand{\TNSNe}{TN\,SNe}
\newcommand{\CCSN}{CC\,SN}
\newcommand{\CCSNe}{CC\,SNe}

\newcommand{\Ni}{\ensuremath{^{56}\mbox{Ni}}}
\newcommand{\Msun}{M$_\odot$}

\newcommand{\Av}{{\ensuremath{\mbox{A$_{V}$}}}}
\newcommand{\av}{{\ensuremath{\mbox{A$_{V}$}}}}
\newcommand{\tpk}{{\ensuremath{\mbox{t{$_{\mbox{\tiny pk}}$}}}}}
\newcommand{\mue}{\ensuremath{\mu_{e}}}

\newcommand{\bftheta}{\ensuremath{\mbox{\boldmath$\theta$}}}
\newcommand{\bfPhi}{\ensuremath{{\bf \Phi}}}
\newcommand{\bfD}{\ensuremath{{\bf D}}}

\shorttitle{Fuzzy Supernova Classification}
\shortauthors{Rodney and Tonry}
\slugcomment{accepted for publication in the Astrophysical Journal} 

\begin{document}

\title{Fuzzy Supernova Templates I: Classification}

\author{Steven A. Rodney}
\email{rodney@ifa.hawaii.edu}
\author{John L. Tonry}
\affil{Institute for Astronomy, University of Hawaii\\
	2680 Woodlawn Dr., Honolulu, HI 96822, USA}
\email{jt@ifa.hawaii.edu}

\begin{abstract}
Modern supernova (SN) surveys are now uncovering stellar explosions at
rates that far surpass what the world's spectroscopic resources can
handle.  In order to make full use of these SN datasets, it is
necessary to use analysis methods that depend only on the survey
photometry.  This paper presents two methods for utilizing a set of 
SN light curve templates to classify SN objects. 
In the first case we present an updated version of the Bayesian
Adaptive Template Matching program (BATM).  To address some
shortcomings of that strictly Bayesian approach, we introduce 
a method for Supernova Ontology with Fuzzy Templates (SOFT), which
utilizes Fuzzy Set Theory for the definition and combination of SN
light curve models. 
For well-sampled light curves with a modest signal to noise ratio (S/N$>
10$), the SOFT method can correctly separate thermonuclear (Type
Ia) SNe from core collapse SNe with $\geq$98\% accuracy.  
In addition, the SOFT method has the potential to classify supernovae
into sub-types, providing photometric identification of very rare or
peculiar explosions. 
The accuracy and precision of the SOFT method is verified using Monte 
Carlo simulations as well as real SN light curves from the Sloan
Digital Sky Survey and the SuperNova Legacy Survey.  In a subsequent
paper the SOFT method is extended to address the problem of parameter
estimation, providing estimates of redshift,
distance, and host galaxy extinction without any spectroscopy.

\end{abstract}



Just prior to the turn of the millennium, \citet{Riess:1998} and 
\citet{Perlmutter:1999} used multi-epoch imaging surveys to
show that the expansion of our universe appears to be accelerating.  
This extraordinary result was revealed with
the careful analysis of just a few dozen Type Ia Supernovae (SNIa).  
In recent years, supernova (SN) surveys such as ESSENCE
\citep{Miknaitis:2007,Wood-Vasey:2007} and SNLS
\citep{Howell:2005,Bronder:2008} 
have begun to probe the nature of this Dark Energy with
samples of a few hundred SNIa.  
The next generation of SN surveys $-$ such as
Pan-STARRS\footnote{\url{http://pan-starrs.ifa.hawaii.edu}},
LSST\footnote{\url{http://www.lsst.org}}, and
JDEM\footnote{\url{http://nasascience.nasa.gov/missions/jdem}} $-$ 
will increase the survey yield by another order of magnitude,
producing thousands of new SN detections each year.  With this
dramatic step forward, we will move beyond the saturation point of the
astronomical community: all the time on all the telescopes of the
world will not be enough for spectroscopic follow-up of the newly
discovered SNe.

The success of these modern SN surveys is
strongly dependent on our ability to accurately and precisely evaluate
a SN based on photometric information alone.  
\citet{Poznanski:2002} and \citet{Johnson:2006} have demonstrated that
SN colors can be used to effectively segregate SNe into the 
broad categories of core collapse (CC) and thermonuclear (TN) SNe.  
Once an object has been classified as a SNIa, the shape of the
light curve can be used to determine an accurate luminosity distance 
\citep{Phillips:1993}.  Extensions of this correlation to
more refined parameterizations of the light curve shape
\citep{Hamuy:1996,Goldhaber:2001} and to multi-color light curves
\citep{Jha:2007,Guy:2007} have been increasingly successful at measuring
accurate SNIa luminosity distances.
This combination of photometric classification and parameter
estimation can be executed for rolling SN searches even with only
early photometric data, as in the SNLS survey \citep{Sullivan:2006a}.

The use of exclusively photometric methods for analyzing large SN
datasets raises a number of concerns.  Sample purity is an important
issue, especially because SNe of Type Ib and Ic are difficult to
distinguish from SNIa.  This can lead to significant biases in
SNIa-derived cosmological parameter estimates \citep{Homeier:2005}. 
Also, without a spectroscopic redshift, one must be aware of the
covariance between estimates of redshift and distance.  When the
parameters of interest are poorly constrained by the supernova alone,
one would like to have a mechanism for smoothly integrating
supplementary data, such as host galaxy colors and morphology.

To address these concerns, one branch of light curve evaluation
techniques have turned to the use of Bayesian probability
theory.  \citet{Tonry:2003} utilized the Bayesian Adaptive Template
Matching (BATM) technique (the precursor to the programs presented in
this paper) to determine luminosity distances by comparing observed
SNIa light curves against a library of templates.  
\citet{Kuznetsova:2007} introduced a Bayesian classification scheme
designed to accurately determine SN types even with poorly sampled
light curves, and \citet{Poznanski:2007a}
presented the Automatic Bayesian Classifier, which uses a similar
approach to classify SNe with as little as one epoch of data.

\citet{Barris:2004} outlined one of the major advantages of the
Bayesian approach to SN light curve analysis.  The Bayesian formalism
allows the user to marginalize over any model parameter that is not
relevant to the scientific question being addressed.  For cosmological
studies, one generally wants to compare redshift to luminosity
distance, but for a SN rate study, the redshift may be a 
``nuisance parameter'' that can be integrated away.  In addition, the
Bayesian formalism can provide a complete and accurate treatment of
the uncertainties inherent in an imperfect classification program. 

In this paper we present two related frameworks for classifying SN
light curves using templates with a fixed light curve shape.  
The first of these is a probabilistic, Bayesian approach,
and the second is based on Fuzzy Set Theory. This paper is divided
into 8 sections as follows:  

\begin{itemize}
  \item[\S\ref{sec:ModelParameters}] considers the necessary elements
    of a SN light curve model, and describes how the template-based
    methods presented here are distinguished from other light curve
    fitters by avoiding a parameterization of the light curve shape.  
  \item[\S\ref{sec:TemplateLibrary}] explains the construction of a
    library of template SNe using photometric and spectroscopic data.
  \item[\S\ref{sec:BayesianProbabilities}] presents BATM, the Bayesian
    Adaptive Template Matching method.  
  \item[\S\ref{sec:FuzzyModels}] discusses limitations of the Bayesian
    approach, and introduces SOFT, the Supernova Ontology with Fuzzy
    Templates program.  
  \item[\S\ref{sec:VerificationTests}] describes the use of 
    Monte Carlo simulations and archival SN light curves to verify the
    accuracy of these methods. 
  \item[\S\ref{sec:OtherModels}] considers how additional models can
    be introduced to account for non-SN contaminants and to quantify
    classification uncertainty. 
  \item[\S\ref{sec:Conclusions}] contains a final summary and
    conclusions. 
\end{itemize}

\section{Model Parameters}
\label{sec:ModelParameters}

When we observe a SN light curve, we are collecting information about
the flux as a function of wavelength and time,
$f_{obs}(\lambda,t)$.  If we were to make a list of all the relevant
parameters that influence the observed flux, we could divide
them up into a set of ``physical parameters'' \bfPhi\ and a set of
``location parameters'' \bftheta.  For \TNSNe the physical
parameter set would include things such as the total mass of ejecta ,
the mass of $^{56}$Ni, the metallicity of the progenitor, the type of
companion star, the asphericity of the explosion, etc.:   

$$ {\bf \Phi} = (M_{ej}, M_{Ni}, Z, ...) $$

\noindent The \bfPhi\ vector could be broadened to include \CCSNe as
well by including characteristics such as the type of stellar explosion,
the presence of Hydrogen or Helium envelopes, etc.
If we had a complete model for all SN progenitor systems and
a perfect SN explosion simulator, then we could define a translation
function $T$ that takes as input a parameter vector \bfPhi\ and
produces a function describing the intrinsic SN luminosity for any
wavelength and time:  

\begin{equation}
 L(\lambda,t) = T({\bf \Phi},\lambda,t) 
\label{eqn:L(lambda,t)}
\end{equation}


For both TN and CC SNe, the object's location in time and
space distorts this intrinsic luminosity in a predictable way, and we
observe the flux.  This distortion is predictable because we have
reasonably good models describing the way SN light is affected by
cosmological distances, dust, and time, so we can write the observable
flux as: 

\begin{equation}
  f(\lambda,t) = \frac{1}{4\pi d_L^2} ~
  L\left(\frac{\lambda}{1+z},t-t_{pk}\right) ~
  10^{-0.4 A_{\lambda/(1+z)}} 
  \label{eqn:f(lambda,t)}
\end{equation}

\noindent where z is the redshift, $d_L$ is the luminosity distance, 
 $A_{\lambda/(1+z)}$ describes the host galaxy extinction, and
\tpk\ is the time of the light curve peak. Hereafter, we will drop the
$\lambda$ dependency when discussing fluxes, under the requirement
that flux comparisons are always to be done with fluxes from the same
broad-band filter. Throughout this paper, we
use a simplified formulation of the four ``location parameters:''

$$ \bftheta = (z, \mue, \Av, \tpk) $$

\noindent 
The first parameter is the redshift of the object (z), which
determines what portion of the spectral energy distribution (SED) is
observed in each bandpass, as well as setting the time-scale of the
light curve due to cosmological time dilation.  
For the second parameter we have recast the luminosity distance using
the parameter \mue.  We define \mue\ as the difference between the
true distance modulus at a given redshift and the value expected for
an empty universe at that z:\footnote{
Note that this formulation for the luminosity distance does not mean
that BATM and SOFT are dependent on an assumed cosmology.  
In Equation~\ref{eqn:mue} the distance parameter, \mue, quantifies
the dimming (or brightening) of a SN relative to how it would appear
in an empty universe.  The only cosmological parameter appearing here
is $H_0$, which factors out as a constant term, independent of
redshift z. Therefore, when comparing the BATM/SOFT distance estimates
of two SNe at different redshifts, there is no bias introduced by
assuming a cosmology, so long as we restrict ourselves to relative
distance estimates. }  

\begin{equation}
 \mu_e = (m-M) - 5 \log_{10} \left(\frac{cz}{H_0}\left(1+\frac{z}{2}\right)\right) - 25
 \label{eqn:mue}
\end{equation}

\noindent The third parameter is the time of the light curve peak
(\tpk), which simply shifts the light curve along the time axis. 
Our final parameter represents the host galaxy extinction (\Av), and
is discussed in detail in \S\ref{sec:HostGalaxyExtinction}.  The
parameter ``\Av'' here is a proxy for a potentially complex extinction
function that may vary with time and wavelength.  Throughout this
paper, we treat \bftheta\ as a set of nuisance parameters, to be
marginalized over for the purpose of getting a broad classification
into one of the established SN categories.   The task of extracting
estimates of some parameters of interest (i.e. measuring z and
\mue\ for cosmology) is addressed in the companion paper. 

\subsection{Light Curve Shape}

The core function of any SN light curve evaluation program is to
predict the observable flux from various SN types, over an (infinite)
range of possible location parameters \bftheta.  By comparing the
model to observations we can make statements of probability regarding
a candidate SN's class and location.  So how does one define the model
for predicting light curve shape?

The fundamental problem is that there is a hidden space of physical
parameters \bfPhi\ that determine SN light curve shape, and we do not
have a theoretical model that can provide the translation of
\bfPhi\ into luminosity as in Eq.~\ref{eqn:L(lambda,t)}.
In parameterized SN light curve fitters such as MLCS, SALT, and
SiFTO, this problem is addressed by making the assumption that a set
of shape parameters $\Delta$ can be used as a proxy for the
unobservable \bfPhi\ vector.\footnote{ Here $\Delta$ represents 
  a vector including a width parameter such as $\Delta m_{15}$ or
  ``stretch,''  and possibly also a color parameter and other higher
  order terms.} 
Parametric fitters thus produce a single model $M(\Delta)$,
which they assert is capable of representing any possible physical
parameter set \bfPhi\ by varying the parameter(s) $\Delta$ to modify
the shape of the light curve.
In general, the $M(\Delta)$ model is restricted to TN\,SNe, although
in principle a model for CC\,SNe could be applied, although it would
be more complex and less informative. 

For BATM and SOFT, we take an alternative approach, in which we do not
parameterize the light curve shape in order to represent all
(TN)SNe with a single model.  Instead, we use a finite set of discrete
models, each one representing the shape of a single well-observed SN
from the local universe.   Whereas the parametric approach essentially
tries to fit an empirical function across the hidden \bfPhi\ space,
with BATM we are sampling the \bfPhi\ space at discrete locations.

\section{The Template Library}
\label{sec:TemplateLibrary}

To build up our library of SN light curve models for comparison
against SN candidates there are three principal stages. The first
step is the collection of a suitable range of light curves for all SN
types. Second, we must fit a smooth interpolating function through the 
observed data. Finally, we use a spectral model, constrained by the
photometry, to predict how each template SNe would look if they
appeared at a different location \bftheta.  

\subsection{Low-z SN Models}
\label{sec:LowzSNModels}

We begin by 
collecting light curves of low redshift (z$\lesssim$0.1) SNe that are
well-sampled with data in all five of the standard Johnson-Cousins 
bandpasses: UBVRI. We place particular 
importance on U and B-band data, because these bands are redshifted
into optical and near-infrared wavelengths at z$\gtrsim$0.5.  We also
prefer templates with early time photometry to constrain the rise
time of the light curve.  To avoid introducing biases due to uncertain
distances, wherever possible we restrict our library to SNe that are
in the Hubble flow ($cz~>~2500$~km~s$^{-1}$) or have an independent
distance determination from Cepheids or surface brightness
fluctuations.  The complete set of light curve templates
are listed in Table~\ref{tab:templib}.  

\begin{deluxetable}{@{\hspace{25mm}}lll@{\hspace{25mm}}lll}
  \tablewidth{\textwidth}
  \tablecaption{{Template Light Curves}}
  \tablecolumns{6}
  \tablehead{
    \multicolumn{3}{c}{\bf Thermonuclear} & \multicolumn{3}{c}{\bf Core Collapse} \\
    \multicolumn{1}{r}{Type} & \colhead{SN} & \multicolumn{1}{l}{References\hspace{15mm}} & \colhead{Type} & \colhead{SN} & \colhead{References}
  }

  \startdata
  Ia$^+$  & \object[SN 1990N]{1990N}    &  \ref{ref:Leibundgut:1991}, \ref{ref:Lira:1998}                                                 & Ib/c   & \object[SN 1994I]{1994I}   & \ref{ref:Richmond:1996}, \ref{ref:Tsvetkov:1995}, \ref{ref:Yokoo:1994} \\       
  & \object[SN 1991T]{1991T}    &  \ref{ref:Lira:1998}, \ref{ref:Altavilla:2004}                                                          & & \object[SN 1998bw]{1998bw} & \ref{ref:Sollerman:2000}, \ref{ref:McKenzie:1999}, \ref{ref:Galama:1998} \\            
  & \object[SN 1999aa]{1999aa}  &  \ref{ref:Jha:2002}, \ref{ref:Krisciunas:2000}, \ref{ref:Altavilla:2004}                                & & \object[SN 1999ex]{1999ex} & \ref{ref:Stritzinger:2002}, \ref{ref:Hamuy:2002} \\                                    
  & \object[SN 1999dq]{1999dq}  &  \ref{ref:Jha:2002}                                                                                     & & \object[SN 2002ap]{2002ap} & \ref{ref:Foley:2003}, \ref{ref:Gal-Yam:2002}, \ref{ref:Yoshii:2003} \\          
  & \object[SN 1999gp]{1999gp}  &  \ref{ref:Jha:2002}, \ref{ref:Krisciunas:2001},                                                         & & \object[SN 2004aw]{2004aw} & \ref{ref:Taubenberger:2006} \\                                                         
  & \object[SN 2000cx]{2000cx}  &  \ref{ref:Candia:2003}, \ref{ref:Li:2001a}, \ref{ref:Jha:2002}, \ref{ref:Altavilla:2004}                & IIb    & \object[SN 1993J]{1993J}   & \ref{ref:Lewis:1994}, \ref{ref:Richmond:1994}, \ref{ref:Benson:1994} \\         
  Ia   & \object[SN 1981B]{1981B}    &  \ref{ref:Buta:1983}, \ref{ref:Tsvetkov:1982}                                                      & & \object[SN 1996cb]{1996cb} & \ref{ref:Qiu:1999} \\                                                                  
  & \object[SN 1989B]{1989B}    &  \ref{ref:Wells:1994}                                                                                   & & \object[SN 2008ax]{2008ax} & \ref{ref:Pastorello:2008} \\                                                           
  & \object[SN 1994D]{1994D}    &  \ref{ref:Richmond:1995}, \ref{ref:Patat:1996}, \ref{ref:Meikle:1996}, \ref{ref:Altavilla:2004}         & IIn    & \object[SN 1999el]{1999el} & \ref{ref:Di-Carlo:2002} \\                                                      
  & \object[SN 1996X]{1996X}    &  \ref{ref:Riess:1999b}, \ref{ref:Salvo:2001}                                                            & & \object[SN 1994Y]{1994Y}   & \ref{ref:Ho:2001} \\                                                                   
  & \object[SN 1998ab]{1998ab}  &  \ref{ref:Jha:2002}                                                                                     & & \object[SN 1998S]{1998S}   & \ref{ref:Liu:2000}, \ref{ref:Fassia:2000} \\                                           
  & \object[SN 1998aq]{1998aq}  &  \ref{ref:Riess:2005}                                                                                   & IIL    & \object[SN 1979C]{1979C}   & \ref{ref:deVaucouleurs:1981} \\                                                 
  & \object[SN 1998bu]{1998bu}  &  \ref{ref:Jha:1999}, \ref{ref:Suntzeff:1999}                                                            & & \object[SN 1980K]{1980K}   & \ref{ref:Buta:1982} \\                                                                 
  & \object[SN 1999ee]{1999ee}  &  \ref{ref:Stritzinger:2002}                                                                             & IIP    & \object[SN 1999em]{1999em} & \ref{ref:Leonard:2002a}, \ref{ref:Elmhamdi:2003} \\                             
  & \object[SN 2000ca]{2000ca}  &  \ref{ref:Krisciunas:2004}                                                                              & & \object[SN 1999gi]{1999gi} & \ref{ref:Leonard:2002} \\                                                              
  & \object[SN 2000cn]{2000cn}  &  \ref{ref:Jha:2002}                                                                                     & & \object[SN 2004et]{2004et} & \ref{ref:Misra:2007} \\
  & \object[SN 2000dk]{2000dk}  &  \ref{ref:Jha:2002} \\
  & \object[SN 2001V]{2001V}    &  \ref{ref:Vinko:2003}, \ref{ref:Lair:2006} \\
  & \object[SN 2002bo]{2002bo}  &  \ref{ref:Benetti:2004}, \ref{ref:Szabo:2003}, \ref{ref:Krisciunas:2004} \\
  & \object[SN 2002er]{2002er}  &  \ref{ref:Pignata:2004} \\
  & \object[SN 2005cf]{2005cf}  &  \ref{ref:Wang:2009} \\
  Ia$^-$   & \object[SN 1998bp]{1998bp}  &  \ref{ref:Jha:2002} \\
  & \object[SN 1998de]{1998de}  &  \ref{ref:Jha:2002}, \ref{ref:Modjaz:2001} \\
  & \object[SN 1999by]{1999by}  &  \ref{ref:Garnavich:2004} \\
  & \object[SN 2002cx]{2002cx}  &  \ref{ref:Li:2003} \\
  \enddata
  \vspace{-5mm}
      \tablerefs{
        \begin{footnotesize}
        \begin{inparaenum}[(1)]
          {\item \label{ref:Altavilla:2004}\protect\citealt{Altavilla:2004}\ }
          {\item \label{ref:Benetti:2004}\protect\citealt{Benetti:2004}\ }
          {\item \label{ref:Benson:1994}\protect\citealt{Benson:1994}\ }
          {\item \label{ref:Buta:1982}\protect\citealt{Buta:1982}\ }
          {\item \label{ref:Buta:1983}\protect\citealt{Buta:1983}\ }
          {\item \label{ref:Candia:2003}\protect\citealt{Candia:2003}\ }
          {\item \label{ref:Di-Carlo:2002}\protect\citealt{Di-Carlo:2002}\ }
          {\item \label{ref:Elmhamdi:2003}\protect\citealt{Elmhamdi:2003}\ }
          {\item \label{ref:Fassia:2000}\protect\citealt{Fassia:2000}\ }
          {\item \label{ref:Foley:2003}\protect\citealt{Foley:2003}\ }
          {\item \label{ref:Gal-Yam:2002}\protect\citealt{Gal-Yam:2002}\ }
          {\item \label{ref:Galama:1998}\protect\citealt{Galama:1998}\ }
          {\item \label{ref:Garnavich:2004}\protect\citealt{Garnavich:2004}\ }
          {\item \label{ref:Hamuy:2002}\protect\citealt{Hamuy:2002}\ }
          {\item \label{ref:Ho:2001}\protect\citealt{Ho:2001}\ }
          {\item \label{ref:Jha:1999}\protect\citealt{Jha:1999}\ }
          {\item \label{ref:Jha:2002}\protect\citealt{Jha:2002}\ }
          {\item \label{ref:Krisciunas:2000}\protect\citealt{Krisciunas:2000}\ }
          {\item \label{ref:Krisciunas:2001}\protect\citealt{Krisciunas:2001}\ }
          {\item \label{ref:Krisciunas:2004}\protect\citealt{Krisciunas:2004}\ }
          {\item \label{ref:Lair:2006}\protect\citealt{Lair:2006}\ }
          {\item \label{ref:Leibundgut:1991}\protect\citealt{Leibundgut:1991}\ }
          {\item \label{ref:Leonard:2002}\protect\citealt{Leonard:2002}\ }
          {\item \label{ref:Leonard:2002a}\protect\citealt{Leonard:2002a}\ }
          {\item \label{ref:Lewis:1994}\protect\citealt{Lewis:1994}\ }
          {\item \label{ref:Li:2001a}\protect\citealt{Li:2001a}\ }
          {\item \label{ref:Li:2003}\protect\citealt{Li:2003}\ }
          {\item \label{ref:Lira:1998}\protect\citealt{Lira:1998}\ }
          {\item \label{ref:Liu:2000}\protect\citealt{Liu:2000}\ }
          {\item \label{ref:McKenzie:1999}\protect\citealt{McKenzie:1999}\ }
          {\item \label{ref:Meikle:1996}\protect\citealt{Meikle:1996}\ }
          {\item \label{ref:Misra:2007}\protect\citealt{Misra:2007}\ }
          {\item \label{ref:Modjaz:2001}\protect\citealt{Modjaz:2001}\ }
          {\item \label{ref:Pastorello:2008}\protect\citealt{Pastorello:2008}\ }
          {\item \label{ref:Patat:1996}\protect\citealt{Patat:1996}\ }
          {\item \label{ref:Pignata:2004}\protect\citealt{Pignata:2004}\ }
          {\item \label{ref:Qiu:1999}\protect\citealt{Qiu:1999}\ }
          {\item \label{ref:Richmond:1994}\protect\citealt{Richmond:1994}\ }
          {\item \label{ref:Richmond:1995}\protect\citealt{Richmond:1995}\ }
          {\item \label{ref:Richmond:1996}\protect\citealt{Richmond:1996}\ }
          {\item \label{ref:Riess:1999b}\protect\citealt{Riess:1999b}\ }
          {\item \label{ref:Riess:2005}\protect\citealt{Riess:2005}\ }
          {\item \label{ref:Salvo:2001}\protect\citealt{Salvo:2001}\ }
          {\item \label{ref:Sollerman:2000}\protect\citealt{Sollerman:2000}\ }
          {\item \label{ref:Stritzinger:2002}\protect\citealt{Stritzinger:2002}\ }
          {\item \label{ref:Suntzeff:1999}\protect\citealt{Suntzeff:1999}\ }
          {\item \label{ref:Szabo:2003}\protect\citealt{Szabo:2003}\ }
          {\item \label{ref:Taubenberger:2006}\protect\citealt{Taubenberger:2006}\ }
          {\item \label{ref:Tsvetkov:1982}\protect\citealt{Tsvetkov:1982}\ }
          {\item \label{ref:Tsvetkov:1995}\protect\citealt{Tsvetkov:1995}\ }
          {\item \label{ref:Vinko:2003}\protect\citealt{Vinko:2003}\ }
          {\item \label{ref:Wang:2009}\protect\citealt{Wang:2009}\ }
          {\item \label{ref:Wells:1994}\protect\citealt{Wells:1994}\ }
          {\item \label{ref:Yokoo:1994}\protect\citealt{Yokoo:1994}\ }
          {\item \label{ref:Yoshii:2003}\protect\citealt{Yoshii:2003}\ }
          {\item \label{ref:deVaucouleurs:1981}\protect\citealt{deVaucouleurs:1981}\ }
        \end{inparaenum}
        \end{footnotesize}
      }
  \label{tab:templib}
\end{deluxetable}

Because of their utility for cosmological studies, the literature is
replete with examples of well-sampled TN SN light curves. Most of the
\TNSN\ light curves in Table~\ref{tab:templib} were acquired from the
collection of \citet{Jha:2007}, and their original references are
listed in the table.  For host galaxy extinctions and
distance estimates we have used the compilation of
\citet{Reindl:2005}, or the listed reference. 
We have divided the \TNSN\ into three sub-classes.  The so-called
``Branch-normal'' Type Ia SNe are listed as simply ``Ia.''
Overluminous SNe are grouped into the ``Ia$^+$'' sub-class. This
includes objects that have been labeled as SN~1991T-like and
SN~1999aa-like, as well as the unique SN~2000cx.  The ``Ia$^-$''
sub-class includes underluminous objects of the SN~1991bg-like
variety, as well as the very peculiar SN~2002cx. One of the strengths
of the template-based approach presented here is the ability to 
seamlessly integrate very peculiar objects such as SN~00cx and 02cx
into the light curve fitting procedure.  

For \CCSNe, the pool of quality templates is distinctly shallower.   
Table~\ref{tab:templib} lists the light curve templates for five
\CCSN\ subclasses.
Each CC type is a very heterogeneous set in
terms of peak brightness, light curve shape and colors. With this in
mind, we don't pretend that just a handful of \CCSN\ light curve
templates can be truly representative of an entire class.
Nevertheless, the
collection listed in Table~\ref{tab:templib} is sufficient 
for the modest goal of distinguishing between the CC and TN classes.  
Type Ib and Ic objects are the most likely to be misclassified,
because of their photometric similarity to Type Ia SNe. Our template
library includes three examples of typical Ib/c SNe, as well as two
of the very high energy ``hypernova'' type (SN~1998bw and SN~2002ap).


\begin{figure}[tb]
\centering
\ifrgb 
  \includegraphics[width=\columnwidth]{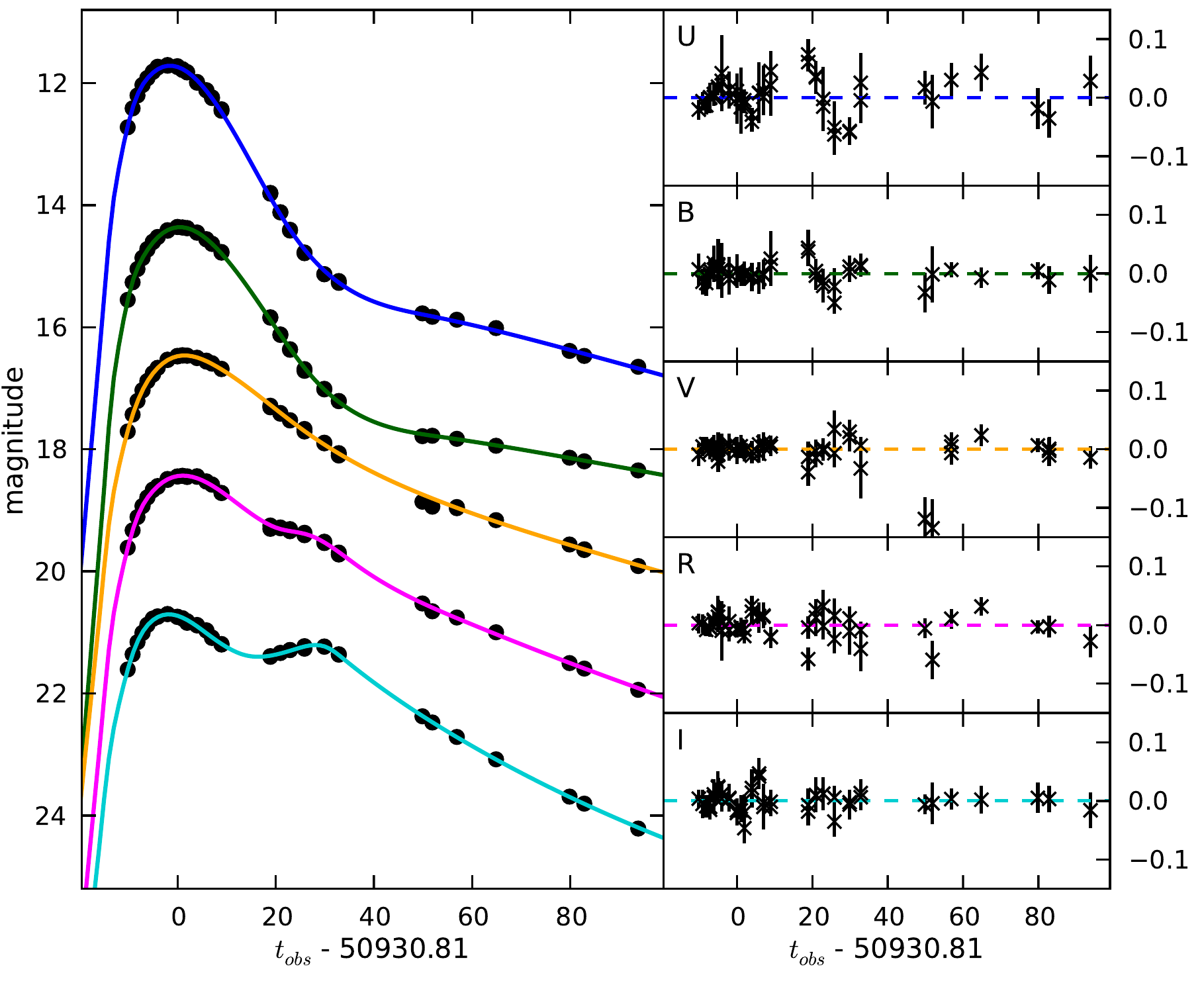}
\else 
  \includegraphics[width=\columnwidth]{\figdir/f01_gray}
\fi
\caption{
  Template spline fits to UBVRI light curves for SN1998aq are
  shown on the left side. Light curve points are shown as crosses,
  and the spline curve is given as a solid line. The light curves for 
  each filter have been  shifted in magnitude for visibility. 
  Residual plots in the panels on the right side show the magnitude
  difference between the observed light curve points and the spline
  fit. The very small and largely uncorrelated residuals
  demonstrate that interpolation with this fit can accurately
  reproduce the true SN colors at times where there is no
  observational data.  
  See the electronic edition of the Journal for a color version of
  this figure. 
}
\label{fig:tempUBVRI}
\end{figure}

With the observed multi-color light curve points in hand, we do not
yet have a functional template, because the discrete observation
points leave large gaps in time where the template flux is as yet
undefined.  To convert the observed data points into a continuous
function we perform a least-squares
fit with a natural cubic spline curve in each bandpass.  
These UBVRI spline curve fits provide
excellent interpolation of the true light curve, as shown in
Figure~\ref{fig:tempUBVRI} for the SN1998aq template.  The residuals
of the observations around the spline fits are consistent with the
observational errors.  For any point
in time, the template spline fits provide UBVRI magnitudes with
uncertainties on the order of 0.02 mags or less.  

\subsection[Redshifting]{Redshifting and Reverse K-correction}

\begin{figure}[tb]
\centering
\ifrgb 
  \includegraphics[width=\columnwidth]{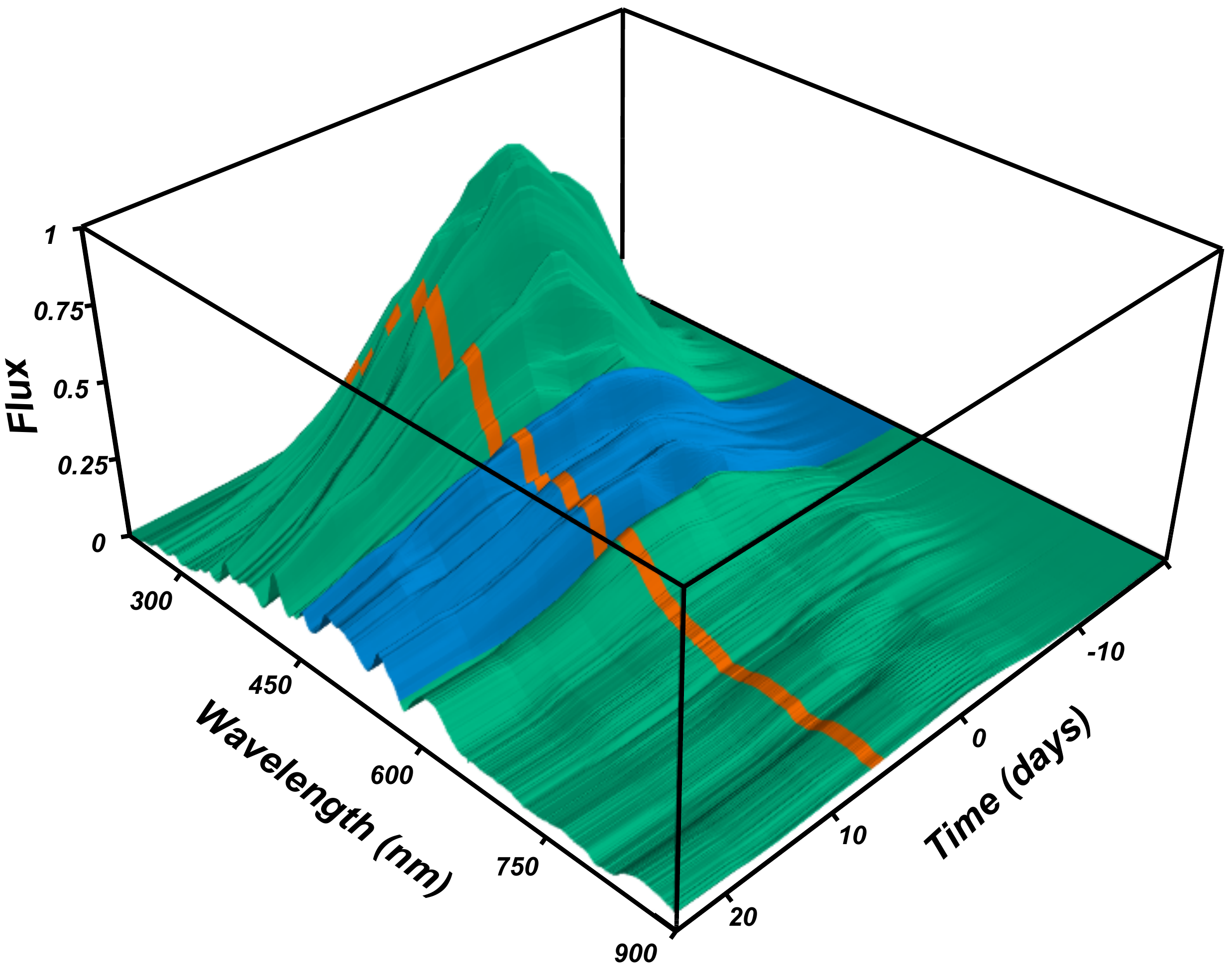}
\else
  \includegraphics[width=\columnwidth]{\figdir/f02_gray}
\fi
\caption{In the ideal scenario for a template library, each template 
would have complete spectrophotometric information, providing the
relative flux value at any wavelength and any time. The thin light
gray ribbon (orange in the electronic version) traces a slice across
the time axis, which provides a single spectrum from t=5 days past
peak.  The broad dark gray (blue) slice indicates the V bandpass.
Integrating  
over the region where the t=5 ribbon overlaps the broad V band would
provide a single data point for a template light curve.
See the electronic edition of the Journal for a color version of
this figure. 
}
\label{fig:snIa_spectra}
\end{figure}

We would now like to adapt these low-z UBVRI light curves to a grid
of points in the four  dimensions of location parameter space 
\bftheta = (z, \mue, \tpk, \Av), while simultaneously 
translating them into the filters used for observing the candidate
object (for this paper we have used $griz$ filters from the Pan-STARRS
survey for our Monte Carlo simulations, plus the 
$g'r'i'z'$ filter set from SDSS). 
To carry out this process 
perfectly, we would need to have a complete  spectrum of the template
object at all points in time (Figure~\ref{fig:snIa_spectra}).  This
ideal situation is unattainable, so we must approximate it by
collecting spectra from many similar objects and piecing them
together. For the \TNSNe\ we use the composite spectra of
\citet{Nugent:2002}\footnote{\url{http://supernova.lbl.gov/~nugent/nugent\_templates.html}}
and \citet{Hsiao:2007}.\footnote{\url{http://www.astro.uvic.ca/~hsiao}} 
For \CCSNe\ we use the collection of spectra employed by
\citet{Blondin:2007}.\footnote{\url{http://cfa-www.harvard.edu/~sblondin/software/snid}}

These spectral compilations provide a complete SED at a set of
discrete points in time, but we want to be able to provide a template
magnitude at {\em any} point in time. To do this, we can use our
temporally complete photometric model (i.e. Figure
\ref{fig:tempUBVRI}) to constrain the time evolution of our sparsely
sampled spectral model in four steps:
\begin{enumerate}
\item  Correct the SED for extinction.
\item  Iteratively warp the SED until the integrated broad-band
  fluxes match the UBVRI flux values drawn from 
  the template spline at the time the SED was taken.
\item  Redshift the warped SED to the desired z.
\item  Integrate the SED over the bandpass of interest to get a
  new broad-band magnitude. 
\end{enumerate}

This process is repeated for every available SED, producing
a new set of multicolor light curve points that represents what the
input template would look like if it appeared at the given redshift z
and were observed in the given bandpass.  At the end of this process,
we have effectively translated the UBVRI input data from our templates
in Table~\ref{tab:templib} into $griz$ points at any redshift, using
the SED's as the bridge between.

\begin{figure}[tb]
\centering
\ifrgb 
  \includegraphics[width=\columnwidth]{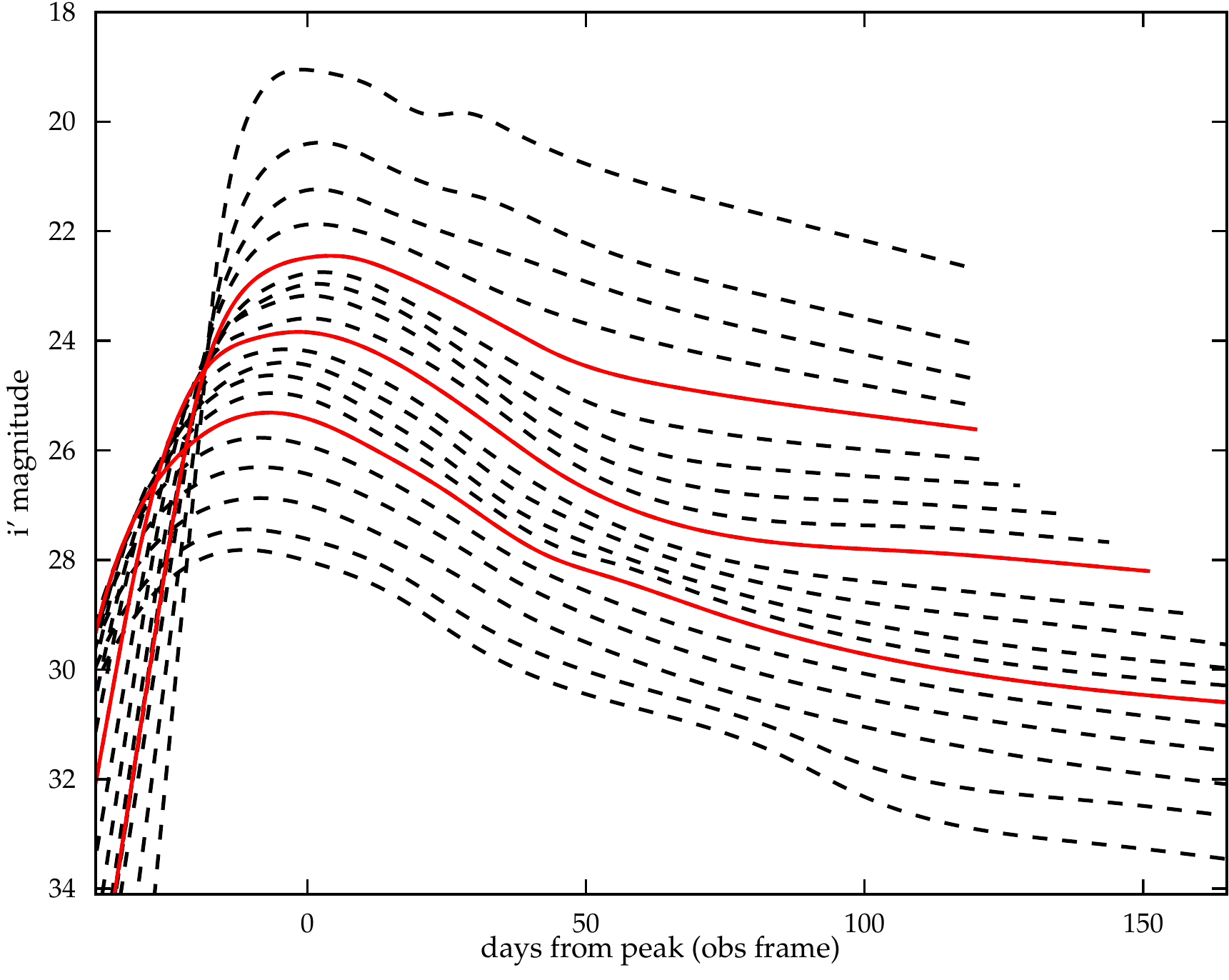}
\else
  \includegraphics[width=\columnwidth]{\figdir/f03_gray}
\fi
\caption{ A subset of the family of redshifted light curve templates
  derived from the SN1998aq parent light curves shown in
  Fig.~\ref{fig:tempUBVRI} and the SEDs shown in
  Fig.~\ref{fig:snIa_spectra}. All light curves are in the PS1 i-band
  with no shifts applied for extinction.  The light curves are spaced
  in redshift from z=0.1 to 1.5 with steps of 0.1, with highlighted
  (red) curves at z=0.5, 1.0, and 1.5. Each light curve has been
  dimmed appropriately for its redshift, if it were in an empty
  universe.  See the electronic edition of the Journal for a color
  version of this figure.  }
\label{fig:splinez}
\end{figure}

Now each of these high-redshift $griz$ light curve models has a
discrete point in each bandpass for every one of the 60+ spectral
templates.  The dates of observation for
the candidate light curves will not coincide with the times of these
discrete points, so we once again turn to spline
interpolation to fill in the gaps.  The same process of least-squares
spline fitting that was used for the low-z UBVRI light curves is now
carried out for each redshifted $griz$ light curve. 
Figure~\ref{fig:splinez} shows the set of spline curves created from
warping the SN1998aq template to 15 redshifts. These
resulting spline curve templates are ``adaptive'' in that they can
be shifted up or down to accommodate a different value of \mue,
reddened to account for host \Av, and
shifted left or right for different values of \tpk.  

Note that when applying a redshift in step 3 we avoid making any
assumptions of cosmological parameters by applying the redshift as if
the template SN were in an empty universe 
($\Omega_M =\Omega_{\lambda} = 0$).  Because we have also defined the
distance modulus \mue\ relative to an empty universe
(Eq.~\ref{eqn:mue}), it becomes a trivial addition operation to adjust
any redshifted template to a new luminosity distance \mue.   

\subsection{Host Galaxy Extinction}
\label{sec:HostGalaxyExtinction}

Dust extinction in a SN host galaxy can mimic the dimming of a
light curve due to distance.  This adds additional scatter -- or
possibly a systematic bias -- to any relations derived from SN
distances. Given that one primary goal of our SN light curve analysis 
program is to enable cosmology with photometry, BATM and SOFT must be
able to disentangle these very similar effects from analysis of the
light curve shape alone.

For an arbitrary bandpass $X$, the host galaxy extinction is given 
by $A_X = ({A_X}/{A_V}) * A_V$ magnitudes.  The visual extinction
\Av\ is a free parameter in our models, so we need to determine the
ratio ${A_X}/{A_V}$ for each bandpass of interest in order to
synthetically redden our template spline curves to match observations. 
As described above, each template light curve is
constructed from a collection of spectra that have been appropriately
redshifted and warped to match the template's photometric light curve. 
As the template library is being built, we integrate each of these 
redshifted template spectra over the bandpass $X$ to get an
unextinguished magnitude $m_0$.  We then apply the R$_V$=3.1
extinction law of \citet{Cardelli:1989} to the spectrum and
repeat the spectral integration to get a reddened magnitude
$m_{red}$. 
The reddening for our bandpass $X$ is now defined as 
$A_X \equiv (m_{red}-m_0)$.  This procedure is repeated for several
input values of \Av\ to get a sampling of $A_X(A_V)$, from which we
can compute the slope ${A_X}/{A_V}$.   Each light curve in the
template library then carries with it a measurement of the 
${A_X}/{A_V}$ ratio, allowing for $A_X$ to be computed very
efficiently for any physically appropriate value of \Av. 

\citet{Nugent:2002} and \citet{Jha:2007}  have measured the time
dependence of extinction in Type Ia SN light curves, demonstrating that
$A_X$ can vary by a few percent over the first few months after
explosion (reaching as much as 5\% in the I band at early times).  This
low-level reddening variation should be included in our handling of
the SN templates, but further work is needed to quantify this effect
on CC SN light curves.  For the present work, we use the median $A_X$
value at all times.  

An additional concern is whether the ratio of total to selective extinction,
R$_V$, may have substantial variability from galaxy to galaxy or SN to
SN.  When the extinction due to dust in the host galaxy of a SN is
low, we can safely ignore the effects of a non-standard extinction law.  
However, any highly reddened SNe with a different host R$_V$
will produce systematically biased parameter estimates.  This can
easily be addressed within the BATM/SOFT framework by allowing R$_V$
to be a free parameter (which can be marginalized over). 
To simplify the Monte Carlo testing presented in this paper, 
all of our simulated light curves are generated with a fixed extinction
law that is set to the Milky Way value of R$_V$=3.1

Furthermore, the extinction law for SNe may change at higher redshift
due to increased rates of star formation, and it is possible to
account for this effect in the choice of prior \citep{Holwerda:2008}.
Additionally, selection effects lead to lower detection rates of
highly extinguished SNe at high z. 
Accordingly, the \Av\ prior should change with z, although in our
verification tests in \S\ref{sec:VerificationTests} we employ a single 
\Av\ prior at all redshifts.   
For testing with synthetic SNe, the \Av\ distribution can be dictated
in the data simulation so the use of an \Av\ prior that is constant
with z is appropriate. A more complex \Av\ prior would be preferable
when applying these methods to high-redshift SN surveys.

\section{Bayesian Probabilities}
\label{sec:BayesianProbabilities}

To develop the probabilistic framework of the BATM method, let us
start with a candidate object called ``SN\,X.''  Suppose we have N
data points in the light curve of SN\,X for which we have measured the
time $t$, flux $f$, and flux uncertainty $\sigma$, so that the
i$^{th}$ data point is given by $D_i=(t_i,f_i,\sigma_i)$.  Each model
$M_j$ at each location \bftheta\ gives a single deterministic
prediction for the observable flux as a function of time:
$\mathcal{F}_{j}(t,\theta)$.  To begin with, we assume that there 
is no uncertainty inherent to the model (we will reexamine this
assumption in \S\ref{sec:FuzzyModels}).  Thus, if the physical model
$M_j$ and location parameters \bftheta\ are correct, then the
observed fluxes of our candidate SN~X should deviate from the model
only by random observational errors.  We assume those errors have a
Gaussian distribution, so the probability of observing the i$^{th}$
data point is just the probability of the error term:

\begin{equation}
  p(D_i|\bftheta,M_j) = \frac{1}{\sqrt{2\pi}\sigma_i} 
  \mbox{exp}\left(\frac{-(f_i-\mathcal{F}_{j}(t_i,\theta))^2}{2\sigma_i^2}\right)
\label{eqn:p(Di|theta,Mj)}
\end{equation}

If the errors are independent, then we can compute the 
likelihood\footnote{
Note that this is in fact a {\em likelihood} and not yet a true
{\em probability}, in the sense that the integral of
Eq.~\ref{eqn:p(D|theta,Mj)} over all models $M_j$ and all
locations \bftheta\ will not be equal to unity. Thus, 
$p(\bfD|M_j,\bftheta)$ is merely a dimensionless function that
quantifies the likelihood that the data set $\bfD$ could be
observed under the assumption that $M_j$ and \bftheta\ are a perfect
model for SN~X. 
}
of a match between SN~X and model $M_j$ at location \bftheta\ as: 

\begin{equation}
   p(\bfD|\bftheta,M_j) = \prod_{i=1}^N p(D_i|\bftheta,M_j)
\label{eqn:p(D|theta,Mj)}
\end{equation}

To determine the net likelihood of model $M_j$, we multiply
Equation~\ref{eqn:p(D|theta,Mj)} by $p(\bftheta|M_j)$ -- the prior
probability distribution  for location parameters (see
\S\ref{sec:Priors}) -- and integrate over all possible vectors 
\bftheta: 

\begin{equation}
p(\bfD|M_j) = \int\limits_{\theta}{ 
 p(\bftheta|M_j)\,p(\bfD|\bftheta,M_j)~d\bftheta}
\label{eqn:p(D|Mj)}
\end{equation}

If we have only one model available, or if we assume that the model
$M_j$ is correct, then we can use Bayes' Theorem to define the
posterior probability as a function of location \bftheta:

\begin{equation}
p(\bftheta|\bfD,M_j) = \frac{ p(\bftheta|M_j)~ p(\bfD|\bftheta,M_j)}
    {p(\bfD|M_j)}
\label{eqn:p(theta|D,Mj)}
\end{equation}

In our case, however, we are considering a list of many applicable
models.  To evaluate the success of each model in reproducing the
data, we must repeat the calculation of Eq.~\ref{eqn:p(D|Mj)} for all
available models $M_j$.  We can then apply 
Bayes' Theorem to define the Posterior Model Probability (PMP), or the
``model weight,'' as: 

\begin{equation}
  PMP_j = p(M_j|\bfD) = \frac{ p(M_j)\,p(\bfD|M_j)}{
    \sum\limits_k~p(M_k)\,p(\bfD|M_k)}
\label{eqn:PMPj}
\end{equation}

Each PMP is a scalar value between 0 and 1 that indicates
the degree to which a given model $M_j$ is preferred by the data 
\bfD, relative to all other models used.  The PMPs from all models
can be combined with posterior probabilities from
Eq.~\ref{eqn:p(theta|D,Mj)} to get model-weighted parameter estimates,
using the method of Bayesian Model Averaging \citep{Hoeting:1999}. The
problem of parameter estimation is addressed in more detail in a
companion paper (Rodney and Tonry, 2009), while for the remainder of
this discussion we restrict ourselves to the task of model selection.

\subsection{Priors}
\label{sec:Priors}

As in all Bayesian applications, the choice of prior probabilities
must be undertaken with some care, to avoid introducing unintended
biases.  For Equations~\ref{eqn:p(D|Mj)} and \ref{eqn:PMPj}
we need to define two types of priors: 
(1) a scalar prior for each model $M_j$, based 
on the expected rate of SNe in each class; 
and (2) a prior probability distribution over \bftheta, reflecting our
initial understanding of the allocation of SNe throughout the universe. 

\subsubsection{Model Priors}
\label{sec:ModelPriors}

The model prior $p(M_j)$ should describe the a priori probability that
a supernova matching the model $M_j$ could be discovered by our survey,
regardless of location \bftheta. We have set the model priors for each
SN sub-class in Table~\ref{tab:templib} according to the following
formula:  

\begin{equation}
p(M_j) = \frac{\mathcal{R}_{class} F_{disc}}{N_{class}} 
\label{eqn:p(Mj)}
\end{equation}

\noindent The first factor, $\mathcal{R}_{class}$ is the rate
of SNe from the sub-class containing model $M_j$, as derived from past
surveys (we have used the compilations of \citet{Blanc:2008} and
\citet{Cappellaro:1999}).  This intrinsic rate is then scaled by
$F_{disc}$, the fraction of objects from this class that would be
detected by our survey, which we derive using Monte Carlo
simulations of survey operations.  Finally, we divide by a
dilution factor, $N_{class}$, to account for the presence of multiple
discrete models representing the same subclass \citep{George:1999}.
Suppose the prior probability for the II-P class as a whole is 
$\mathcal{R}_{IIP}F_{disc}=0.2$, and we have 3 templates in that
subclass.  If we assign each of the II-P templates a model prior of
$p(M_j)=0.2$, then the total prior weight in 
the II-P class is artificially inflated to 0.6.  Thus, to avoid
introducing a bias due to having different numbers of models, we
divide by the number of templates within each class, $N_{class}$. 

\subsubsection{Location Priors}
\label{sec:LocationPriors}
If we assume that the four location parameters 
\bftheta=(z,\mue,\Av,\tpk) are uncorrelated, then we can
recast the location prior as a product of constituent parts:

\begin{equation}
p(\bftheta|M_j) = p(z|M_j) p(\mue|M_j) p(\Av|M_j) p(\tpk) 
\label{eqn:p(theta|Mj)}
\end{equation}

\noindent Note that the \tpk\ prior is independent of the model
choice, because the calendar date of a SN explosion is not influenced
by the type of SN event. All other location priors  could be dependent
on the model or class of models under consideration (e.g. The
prior for host galaxy extinction might be different for CC vs
TN\,SN). 
Whenever possible, these priors should be informed by
spectroscopy and host galaxy photometry.  In the absence of such 
information, a uniform prior may be applied.  In all cases it is
important to scale each prior probability distribution so that it
integrates to unity over the range of values that BATM has access to. 

In the verification tests presented in \S\ref{sec:VerificationTests} we have
employed a uniform prior for for the redshift $p(z|M_j)$, the
cosmological dimming $p(\mue|M_j)$, and the time of light curve peak
$p(\tpk)$.  For the host galaxy extinction prior
$p(\Av|M_j)$ we have used the ``galactic line-of-sight'' (glos) 
prior employed by the ESSENCE team \citep{Wood-Vasey:2007}, which is
based on the host galaxy extinction models of \citet{Hatano:1998}, 
\citet{Commins:2004}, and \citet{Riello:2005}.  The form of this prior
is an exponential plus a truncated Gaussian, expressing the preference
for matches that select a low \Av\ value.

\subsection{Classification}

To extract a classification probability with BATM is now
straightforward.  We simply compute the PMPs for all models, then sum
together the PMPs from all models belonging to the class of interest.
For example, to determine the probability that an object is of
Type~Ia: 

\begin{equation}
p(Ia|D) = \sum\limits_{k\in\{Ia\}} PMP_k
\label{eqn:p(Ia|D)}
\end{equation}

\noindent where the sum is over just those templates
in the Type Ia subset, $\{Ia\}$.
The classification probability for any other class or subclass can be
found in the same way, and these normalized classification
probabilities are related through simple additive statements:

\begin{equation*}
  \begin{array}{ll}
    p(TN|D) = & p(Ia^+|D) + p(Ia|D) + p(Ia^-|D) \\
  \end{array}
\end{equation*}
\begin{equation*}
  \begin{array}{ll}
    p(CC|D) = & p(Ibc|D) + p(IIb|D) + \\
              & p(IIP|D) + p(IIL|D) + p(IIn|D) \\
  \end{array}
\end{equation*}
\begin{equation*}
    p(TN|D) + p(CC|D) = 1
\end{equation*}

\section{Fuzzy Models}
\label{sec:FuzzyModels}

In the preceding section we laid out the BATM framework for
probabilistic comparison between SN light curve data and
template-based models.  The alternative Fuzzy Set-based method
described here (SOFT) is motivated by a careful consideration of the
assumptions required by BATM.  In order to apply the tools of
probability theory in BATM, we invoked  three key assumptions: 

\begin{enumerate}
\item Each SN model $M_j$ provides an exact deterministic prediction
  for the flux as a function of time. 
\item The set of all models $\{M\}$ is complete : all possible light
  curve shapes are represented.
\item The set $\{M\}$ is non-redundant : a candidate SN cannot be a
  true match to more than one $M_j$.
\end{enumerate}

\noindent These assumptions would be justifiable if we were using a
parameterized SN model that ostensibly covers the entire hidden
range of physical parameters \bfPhi.   However, the premise breaks
down when we use a discrete set of templates with fixed light curve
shapes.

In the language of set theory, we can define a universal set $U$
that contains all possible SNe with any physical parameters
\bfPhi\ and any possible location \bftheta.  Each of our discrete set
of models $M_j$ defines a subset $S_j\subset U$ containing all
possible SNe that have precisely the physical parameters \bfPhi$_j$,
observed at any location \bftheta.  These subsets are
classical or ``crisp'' sets, in that membership in the set is a binary
condition: either you are in or you are out. This is in keeping with
the determinism of assumption (1) above, which was used to derive
Eq.s~\ref{eqn:p(Di|theta,Mj)}-\ref{eqn:p(D|theta,Mj)}.  From this we
can immediately see that any {\em discrete} and {\em finite} set of
models cannot satisfy both assumptions (1) and (2): if the models are
presumed to be deterministic, then there will always be some subset
$S_X\subset U$ that is not represented. 

How does this contradiction affect our classification estimates? 
Suppose we have an extremely sparse template library, consisting of 
only two template models, $M_1$ and $M_2$, which have underlying
physical parameters \bfPhi$_1$ and \bfPhi$_2$.\footnote{ 
  This reduction is only used for convenience. The conclusions drawn
  are equally valid for any finite set of models. 
}
Now let us observe a candidate object, SN~X, which has physical
parameters \bfPhi$_X$ that are intermediate between \bfPhi$_1$ and
\bfPhi$_2$ (e.g., perhaps the candidate has a \Ni\ mass that is
bracketed by the two templates: $M_{Ni,1} <  M_{Ni,x} < M_{Ni,2}$).   
SN~X is not a member of either of the crisp subsets $S_1$ and $S_2$
defined by our two models, so the classification likelihood of
Eq.~\ref{eqn:p(D|Mj)} will be very small for both $M_1$ and $M_2$. If
the data quality is high ($\sigma_i \rightarrow 0$), then  the computed
likelihood of a match can become vanishingly small, and we will be
unable to compute a meaningful classification probability.  This is
precisely the opposite of what we expect from a respectable SN
classification algorithm:  high signal to noise observations should
lead to more precise classifications, not more classification
failures. 

The problem we are exposing is a fundamental limitation of
a template-based Bayesian classifier: one cannot reliably classify SNe
that fall into the gaps between discrete templates when constrained by
our three core assumptions. 

\subsection{Defining Fuzzy Sets}
\label{sec:DefiningFuzzySets}

One possible means of addressing this problem of discrete
models is presented by the field of logical analysis known as 
``Fuzzy Set Theory'' \citep{Zadeh:1965}.  To use this mathematical
framework,  we must first adjust our models so that they define
``fuzzy sets,'' whose membership is not crisply defined.  
Some examples of fuzzy sets in the
realm of astronomy are categories such as ``high mass stars'' or
``rich clusters.''  The semantic meaning of these groups is clear
enough, but determining whether a certain object belongs within the
set is somewhat ambiguous.  A 100~\Msun\ star certainly belongs within
the ``high mass'' group, and a 0.5~\Msun\ star does not, but does the
membership line get drawn at 2, 5 or 10 \Msun?  For our purposes, we
want to explicitly introduce some fuzziness into the definition of our
models, so as to ensure that any candidate SN~X has some degree of
membership in at least one of the subsets $S_j$.  As discussed below,
we will need a new method for combining the evidence from
multiple fuzzy models when deriving a composite parameter estimate. 

Membership in a fuzzy set A is defined by a membership function
$g(x|A)$ which takes any possible set member $x$ and assigns it a
``membership grade,'' which is a real number in the range [0,1].  For
crisp sets, the membership function is binary, always providing
$g(x|A)=1$ for set members, or $g(x|A)=0$ for non-members. For fuzzy
sets, a membership grade close to unity indicates a strong association
with the set, while $g(x|A)\sim 0$ suggests the opposite.

To see how we might define a membership function for fuzzy SN sets,
let us consider the family of template light curves derived from 
SN~1998aq, as shown in Figure~\ref{fig:splinez}.  The actual supernova 
event 1998aq had some unknown set of particular physical parameters
\bfPhi$_{98aq}$.  The model we constructed from the 1998aq light curve
template allows us to produce the precise light curves of
Figure~\ref{fig:splinez}, which predict how a SN with identical
physical parameters \bfPhi$_{98aq}$ would appear if it were observed at
a different location \bftheta.   Suppose we introduce a small
perturbation of the physical parameters, $\delta\bfPhi$, perhaps by
adding a little more \Ni\ or allowing for a bit more mixing in the
ejecta.  This makes a slightly different explosion, with
physical characteristics now given by  

$$ \bfPhi_{98aq}^{\prime} = \bfPhi_{98aq} + \delta\bfPhi $$

Recalling the underlying assumption that changes in the
physical parameters will be reflected in changes of the light curve
shape, we can assume that the small physical perturbation
$\delta\bfPhi$ results in a small change to the observable flux: 

$$ f^{\prime}(\lambda,t) = f_{98aq}(\lambda,t) + \delta f(\lambda,t) $$

%
%
\noindent This new explosion would fall outside of the crisp set
defined by the SN\,98aq model, but we can include it within an
expanded fuzzy set that is centered on SN\,98aq.  

What form should the membership function for this fuzzy set take? 
In words, the membership function will define the fuzzy set of ``All
SNe with light curves similar to $M_j$ at location \bftheta.''
To quantify {\em similarity} in this statement, we introduce
$\sigma'_j$ as a ``model fuzziness'' term.  The $\sigma'_j$ term
has units of flux, and it defines the width of a Gaussian distribution
around the model $M_j$ that indicates how much flux deviation from the
core model is allowed for members of this fuzzy set.\footnote{
We have chosen the Gaussian function for convenience.  Another form
(such as a Lorentzian or a Voigt Profile) could be equally
appropriate.} 
Consider our candidate SN~X, and for the moment let us ignore
observational errors ($\sigma_i\sim0$ for all N data points).
If we assume the fuzzy model $M_j$ and location \bftheta\ to
be correct, then the probability of getting the observed flux
discrepancies between the model and the data 
{\em due to model uncertainty alone} is: 

\begin{equation}
   p'(\bfD|\bftheta,M_j) =
   \prod_{i=1}^N{\frac{1}{\sqrt{2\pi}\sigma'_j} 
     \mbox{exp} \left(\frac{-(f_i - \mathcal{F}_{j}(t_i,\theta))^2}{
       2~{\sigma'_j}^2}\right) } 
\label{eqn:p'(D|theta,Mj)}
\end{equation}

This mimics Eq.~\ref{eqn:p(D|theta,Mj)}, which provided the likelihood
of observing a light curve $\bfD$ based on observational errors
only.  To combine these two probability estimates
we convolve Eq.~\ref{eqn:p(D|theta,Mj)} with
Eq.~\ref{eqn:p'(D|theta,Mj)} \citep[e.g.][Sec.4.8]{Gregory:2005}
to get: 

\begin{equation}
  \begin{array}{ll}
  p''(\bfD|\bftheta,M_j) = & 
  \prod\limits_{i=1}^N{\frac{1}{\sqrt{2\pi}\sqrt{ \sigma_i^2 + {\sigma'_j}^2}}}\\
    & \times\ \mbox{exp}\left(\frac{-\left(f_i - \mathcal{F}_{j}(t_i,\theta)\right)^2}{ 2~(\sigma_i^2 + {\sigma'_j}^2)}\right)\\
  \end{array}
\label{eqn:p''(D,theta,Mj)}
\end{equation}

We can then modify this likelihood to account for our priors on
\bftheta\ and $M_j$, which produces a final likelihood estimate
suitable for use as the membership function of a fuzzy set:

\begin{equation}
  g(\bfD|\bftheta,M_j) = p(\bftheta|M_j)~ p(M_j)~ p''(\bfD|\bftheta,M_j)
  \label{eqn:g(D|theta,Mj)}
\end{equation}

\noindent This, then, is the membership function for our candidate SN~X 
in the fuzzy set defined by (\bftheta, $M_j$). This fuzzy set contains
all possible SNe that are similar to the  fuzzy model $M_j$ and are
also at the precise location \bftheta.

These membership functions are very similar to the probabilities used
in \S\ref{sec:BayesianProbabilities}, but there are some important
distinctions.  First, consider that in classical probability theory
when one assumes a given model $M_j$, it is then required that the
posterior probability over \bftheta\ must integrate to unity:

$$ \int\limits_\theta  p(\bftheta|\bfD,M_j)~d\theta = 1 $$

\noindent This requirement is enforced by the presence of the
normalization factor in Eq.~\ref{eqn:p(theta|D,Mj)}.  It is not
necessarily appropriate to apply the same normalization to these fuzzy
membership functions.  

Furthermore, when we apply Bayes' Theorem to compute the PMP in
Eq.~\ref{eqn:PMPj}, the denominator ensures that the sum of all
PMPs is equal to unity:  

$$ \sum\limits_j PMP_j = \sum\limits_j p(M_j|\bfD) =  1 $$

\noindent This normalization is not a feature  of the fuzzy set
membership functions:  

$$ \sum\limits_j \int\limits_\theta  g(\bfD|\bftheta,M_j) d\theta \neq 1 $$

 The reason for this difference is that fuzzy sets can 
overlap, and therefore they do not satisfy the Kolmogorov
axioms.\footnote{Specifically, the 2nd and 3rd Kolmogorov axioms,
  stating that $\sum_i P(e_i)=1$ for a complete set of elementary
  events $e_i$, and that $P(A\cup B\cup C)= P(A) + P(B) + P(C)$.}
In our fuzzy sets a candidate SN can be similar to template
$M_1$ while also being similar to $M_2$.  Thus, any given candidate
can have non-zero membership grades in multiple fuzzy sets, and the
sum of membership grades can be greater than or less than unity.  

\subsection{Combining Fuzzy Sets}
\label{sec:CombiningFuzzySets}

By comparing the light curve of our candidate SN~X against all models
$M_j$, we can derive membership grades $g(\bfD_X|\bftheta,M_j)$ for all
possible values of \bftheta.  
We would now like to combine the results from all the \TNSN\ model
comparisons to get a composite membership grade for the fuzzy set
containing all \TNSNe.  After also computing a composite fuzzy
membership grade for the \CCSN\ class, we could compare the relative
strengths of membership to determine a classification. 
How do we go about combining multiple fuzzy set membership functions
from Eq.~\ref{eqn:g(D|theta,Mj)}?

\citet{Zadeh:1965} in his seminal paper on fuzzy set theory proposed
the following rules defining the ``fuzzy union'' and 
``fuzzy intersection'' of two fuzzy sets: 

\begin{equation}
  \begin{array}{l}
    g(X|{A\cup B}) = \max\left[ g(X|A), g(X|B) \right]\\
    g(X|{A\cap B}) = \min\left[ g(X|A), g(X|B) \right]
    \label{ref:comborules1}
  \end{array}
\end{equation}

\begin{figure}[!tb]
\centering
\ifrgb 
  \includegraphics[width=\columnwidth]{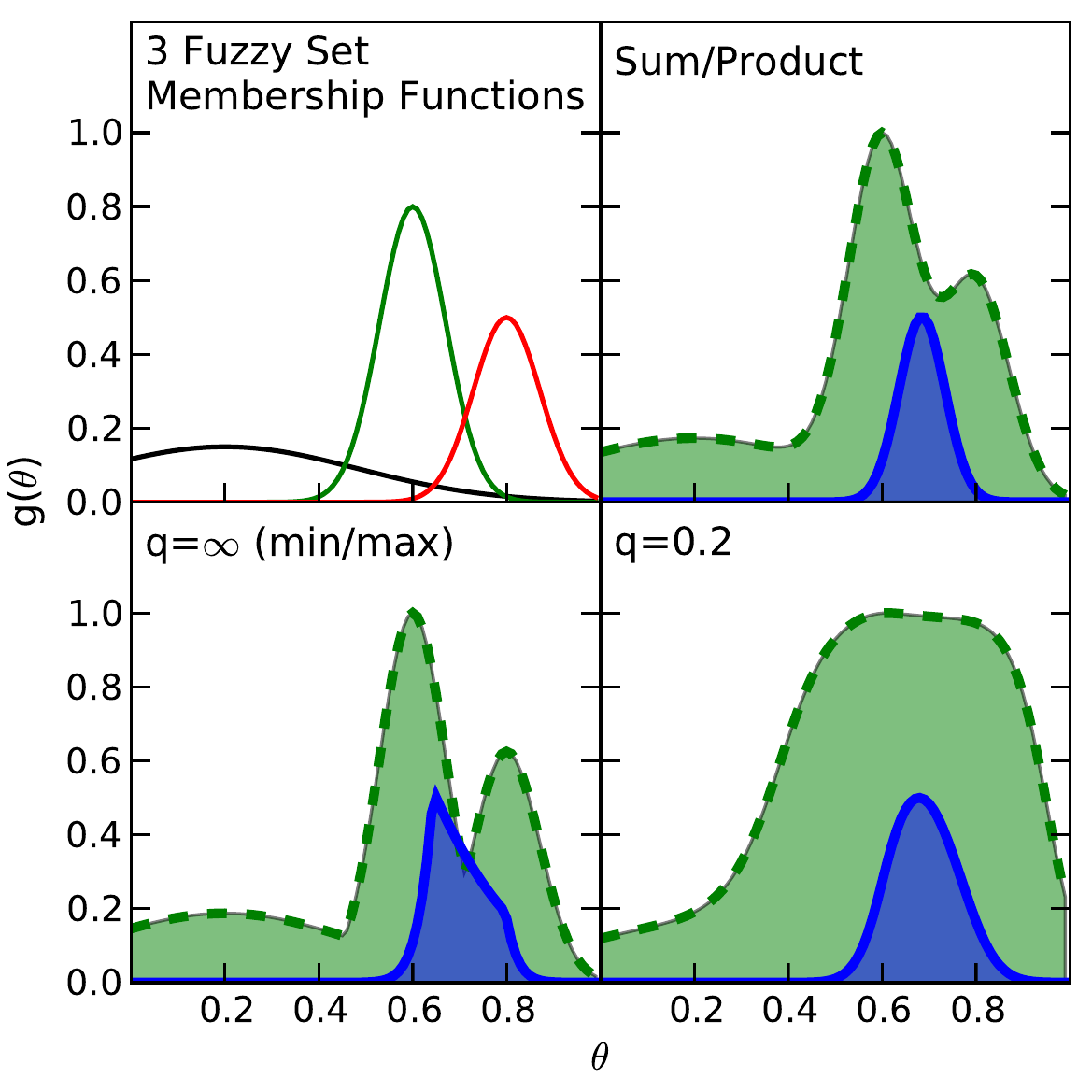}
\else
  \includegraphics[width=\columnwidth]{\figdir/f04_gray}
\fi
\caption{Demonstration of combination rules for fuzzy sets. In the
  upper left panel are shown three hypothetical fuzzy set membership
  functions  $g_1$, $g_2$, and $g_3$.  The remaining three panels show
  different options for the combination of these fuzzy sets. In each
  case, the aggregate fuzzy set created by applying the conjunction
  operator (OR) is shown as a (green) dashed line: 
  $g_1\cup g_2\cup  g_3$.  The disjunction operator (AND) appears as 
  a filled curve (blue), showing $g_1\cap g_2\cap  g_3$ 
  These combined sets are displayed  with arbitrary normalization for
  clarity. 
  {\em Upper Right:} If the membership functions were treated as
  classical probability functions, then one would use the Sum operator
  for OR, with the Product operator for AND.
  {\em Lower Left:} Applying the Dombi functions for fuzzy set
  operators with $q=\infty$, we recover a special case where OR is the
  $\max$ function, and the $min$ function serves as the AND operator.
  {\em Lower Right:} A small value of $q=0.2$ emphasizes agreement
  between fuzzy sets. The resulting union and intersection sets are
  less sharply peaked, and less strongly influenced by outliers.
}
\label{fig:fuzzyOperators}
\end{figure}

\noindent Zadeh's combination rules are illustrated graphically in the
left two panels of
Figure~\ref{fig:fuzzyOperators}. \citet{Bellman:1973} provided an 
axiomatic framework for fuzzy set operators that justifies Zadeh's
min/max rules as the purest possible operators. That is, the max
function provides the most inclusive fuzzy OR, while the min function
is the most exclusive AND operator.  However, the min/max functions
are not the only allowable operators. Several classes of functions
can satisfy the required axioms while also allowing some variation in
the level of inclusiveness or exclusiveness
\citep[e.g.][]{Hamacher:1978,Yager:1980,Dubois:1980a}. 
These operators are called {\em interactive} because rather than
choosing just one of the membership grades $g(X|A)$ or $g(X|B)$
as the min/max operators do, these functions allow both the A and B
functions to contribute simultaneously. For further discussion of
the axiomatic requirements of fuzzy set operators and comparisons of 
some interactive classes, see section II.1 of \citet{Dubois:1980} and
Chapter 2 of \citet{Klir:1988}. 

\citet{Dombi:1982} derived a family of fuzzy set operators that is
general enough to encompass the operators of \citet{Zadeh:1965},
\citet{Hamacher:1978}, and \citet{Yager:1980} as special cases. 
Let $a$ and $b$ represent the membership grades of our candidate
object $X$ for the fuzzy sets $A$ and $B$, respectively.  Then the
Dombi membership function for $X$ in the fuzzy union set $A\cup B$ is: 

\begin{equation}
  \begin{array}{l}
    g(X|{A\cup B}) = \frac{1}{ 1 +
      \left[\left(\frac{1}{a}-1\right)^{-q} + 
        \left(\frac{1}{b}-1\right)^{-q} \right]^{-1/q} }
    \label{eqn:fuzzyOR}
  \end{array}
\end{equation}

\noindent and for the fuzzy intersection $A\cap B$ Dombi has:

\begin{equation}
  \begin{array}{l}
    g(X|{A\cap B}) =\frac{1}{ 1 +
      \left[\left(\frac{1}{a}-1\right)^{q} + 
        \left(\frac{1}{b}-1\right)^{q} \right]^{1/q} }
    \label{eqn:fuzzyAND}
  \end{array}
\end{equation}

\noindent where the parameter $q$ is in the range $(0,\infty)$.
Figure~\ref{fig:fuzzyOperators} shows the shapes of union
and intersection sets resulting from a combination of three fuzzy set
membership functions.
The $q$ parameter governs the strength of interaction between the two
membership function arguments A and B.  In the limit where
q$\rightarrow\infty$ the Dombi union function reduces to Zadeh's $\max$
operator, while the Dombi intersection becomes Zadeh's $\min$
function.  As the value of $q$ decreases, these fuzzy operators
become progressively ``softer,'' allowing more and more interaction
between the sets.

\subsection{Calibrating Fuzzy SN Models}

With the fuzzy SN models and combination operators described above, we
now have a complete mathematical framework for combining fuzzy sets. 
To get here, we have introduced two new parameters: $\sigma'_j$
describes the amount of fuzziness in our SN models, and $q$ controls
the amount of interaction allowed when combining sets.  These
parameters are actually characteristics of the template library, and
may be estimated empirically by inter-comparison of templates.  

If our template library consisted of
just two or three models with substantially different physical
characteristics \bfPhi, then we would need to make them very fuzzy (a
very large $\sigma'_j$) in order to fill the space between models. With
a few hundred templates the need for fuzziness would be reduced, so
$\sigma'_j$ could be very small.   
To calibrate the amount of fuzziness required for our template
library, we measured the flux difference between pairs of templates
belonging to the same SN sub-class, at the same location
\bftheta.  Taking a median value from all the possible pairwise
comparisons, we determined that setting $\sigma'_j$ to be 10\% of the
template flux (approximately 0.1 mag) is an appropriate degree of
fuzziness for the set of \TNSN\ templates listed in
Table~\ref{tab:templib}.  For the set of \CCSNe, with fewer templates
and greater heterogeneity, we found that slightly more fuzziness is
necessary, at the level of 
$\sigma'_j = 0.15 ~ \mathcal{F}_{j}(t,\theta)$.  

As shown in Figure~\ref{fig:fuzzyOperators}, setting $q=0.2$ results
in a fuzzy union (dashed line) that emphasizes the overlap between
sets. In this sense it is a sort of hybrid between the strict
interpretations of the classical OR and AND operators.  Thus, the
$q=0.2$ fuzzy union achieves the goal stated at the top of
\S\ref{sec:CombiningFuzzySets} in that it indicates how much a
candidate SN~X is similar to {\em any or all} of the available models.

A more rigorous empirical calibration of $q$ could be done using a
training set of SNe with known locations \bftheta.  
After comparing the training set against the template library, one
would adjust the $q$ parameter so that fuzzy model combination
provides the most accurate estimate of the true location.  Such a
detailed calibration of $q$ is beyond the scope of this work, but we
have performed preliminary tests using \TNSN\ light curves from the
Sloan Digital Sky Survey (SDSS) \citep{Holtzman:2008}.  This rough,
empirical calibration showed that $q$ values in the range (0.1,1.0)  
provide acceptable results with this template library.

\subsection{Fuzzy Classification}
\label{sec:FuzzyClassification}

We now have all the necessary tools for executing the SOFT method on
SN light curves. The procedure is as follows: 

\begin{enumerate}
\item For each template $M_j$, at every location \bftheta, subtract the
  model flux $\mathcal{F}_j(t,\theta)$ from the observed flux
  of the candidate $f_i(t)$.
\item Using
  Eq.'s~\ref{eqn:p''(D,theta,Mj)}-\ref{eqn:g(D|theta,Mj)},
  determine the fuzzy set membership grade of this candidate over all
  available parameter space.
\item Collect membership functions into groups according to SN
  sub-class, excluding any that are zero everywhere.
\item Combine the membership grades within each sub-class using the
  fuzzy union operator of Eq.~\ref{eqn:fuzzyOR}
\end{enumerate}

The result is a final composite membership function for each SN
sub-class, $g(\bfD|\theta,Y)$, giving the degree of membership for
SN~X in sub-class~Y as a function of location \bftheta.  The peak of
the distribution $g(\bfD|\theta,Y)$ provides an estimate of the true
location \bftheta, and the width of $g(\bfD|\theta,Y)$ indicates the
uncertainty of that estimate (see Paper 2 for a complete discussion of
parameter estimation with fuzzy templates).  The integral of
$g(\bfD|\theta,Y)$ over \bftheta\ gives a classification grade for
that particular sub-class.

Given the similarity between fuzzy set classification grades and
Bayesian posterior model probabilities, one might be tempted to
normalize each sub-class grade by the sum over all sub-classes, in a
manner similar to Eq.~\ref{eqn:PMPj}.  In general this is not
appropriate, because fuzzy sub-classes are not mutually exclusive, and
therefore don't follow the Kolmogorov axiom of additivity: 
$P_{A\cup  B\cup C} = P_A+P_B+P_C$.  Any subclasses that might have
some membership interaction should be combined using Dombi's fuzzy
intersection or union operators.  
The one useful exception arises when comparing relative
classification grades for the TN and CC SN classes.  In that case we
can assert that these two classes represent fully isolated sets,
because there is a significant physical  difference between any Type Ia
and any core collapse SN.  Thus, the TN and CC fuzzy sets can be
treated as crisp sets and combined additively.

The normalized membership grade for the \TNSN\ is analogous to the
Posterior Model Probability (PMP) of Eq.~\ref{eqn:PMPj}.  We call this
quantity the Posterior Membership Grade (PMG), and calculate it as
follows:

\begin{equation}
PMG_{TN} = g(TN|\bfD) = \frac{\bigcup\limits_{i\in TN} g(\bfD|M_i)}{
  \bigcup\limits_{i\in TN} g(\bfD|M_i) + \bigcup\limits_{j\in CC} g(\bfD|M_j)}
\label{eqn:PMGTN}
\end{equation}

\noindent where $\bigcup\limits_{i\in TN} g(\bfD|M_i)$ 
 
indicates the fuzzy
union of membership grades from all templates in the \TNSN\ class. The
corresponding classification grade for \CCSNe, $PMG_{CC}$, is
defined in the same manner.

\section{Verification Tests}
\label{sec:VerificationTests}

To evaluate the utility of the SOFT template matching
technique, we first generated a Monte Carlo simulation containing
5000 synthetic light curves for each of the \TNSN\ and \CCSN\ classes,
evenly distributed over the sub-classes. 
Each synthetic light curves was created using a template that was
prepared in the way described in 
\S\ref{sec:TemplateLibrary}. To begin, the template was warped and  
shifted to a randomly selected point in the bounded 4-dimensional
parameter space of \bftheta=(z, \mue, \Av, \tpk).  We  then generated
an ``observed'' $griz$ light curve for each synthetic object, using a
simulation modeling 120 nights of the Pan-STARRS 1 Medium Deep survey.
In terms of wavelength coverage, cadence, and depth, this simulation is
roughly intermediate between the SDSS-II Supernova Survey
\citep{Sako:2008} and planned surveys from projects such as the Large
Synoptic Survey Telescope \citep{Tyson:2002}.  

The location parameters
\bftheta=(z,\mue,\Av,\tpk) were drawn from a uniform distribution over
a broad range of parameter space: 

\begin{align*}
0.001<z<1.0\\
-0.8<\mbox{\mue}<0.8\\
0.0<\mbox{\Av}<1.5\\
20<\mbox{\tpk}<100\\
\end{align*}

\noindent By choosing a uniform
distribution in z, \mue, and \av, these Monte Carlo simulations
are clearly not depicting a realistic survey yield, where the rate of
detection depends on cosmology and the underlying SN explosion
rate. The purpose here is to determine whether there are regions of
parameter space where the SOFT methods are poor classifiers,
which could lead to sample selection biases.

Our second data set for testing classification accuracies is a
collection of 222 published light curves from the SDSS-II and SNLS
surveys. The SDSS light curves, from \citet{Holtzman:2008} consist of
146 objects that have spectroscopic redshifts ranging from
0.01 to 0.45.   All of these objects
were classified as Type Ia, although 16 were given a ``SNIa?''
classification, indicating some uncertainty in their designation as
\TNSNe.  These SNe are well sampled in the Sloan
$u'g'r'i'z'$ bands, with observations
done in all 5 filters on a single night, a typical cadence of less
than 5 nights, and sensitivities of around 22 mags. 

From the SNLS survey we include 71 objects from \citet{Astier:2006}
with spectroscopic redshifts ranging from 0.25 to 1.0.  
Fifteen objects in this set were given the classification ``Ia*'' to
indicate that the best spectral match is a SN Ia model, but that an
alternative classification cannot be ruled out.  The $griz$ light
curves of these 71 SNe have a similar sampling 
frequency to the SDSS set, with typically 8-15 observational epochs
over 100 days in three or more filters.  Finally, we include an
additional 5 SNLS light curves from \citet{Nugent:2006} that are
classified as Type II-P supernovae.  These objects have redshifts
between 0.1 and 0.3. 

\begin{figure*}[!tb]
\centering
\ifrgb 
  \includegraphics[width=0.45\textwidth]{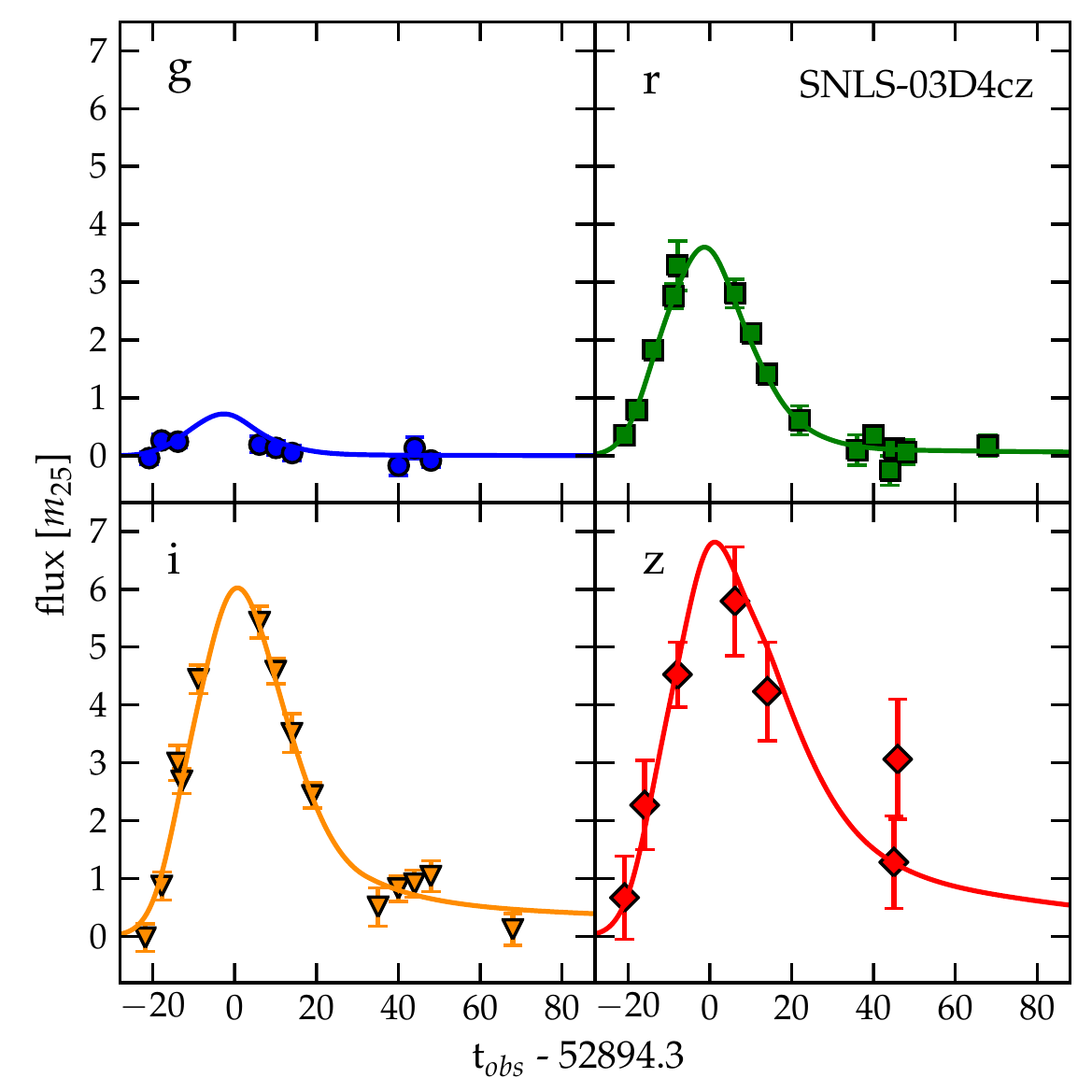}
  \includegraphics[width=0.45\textwidth]{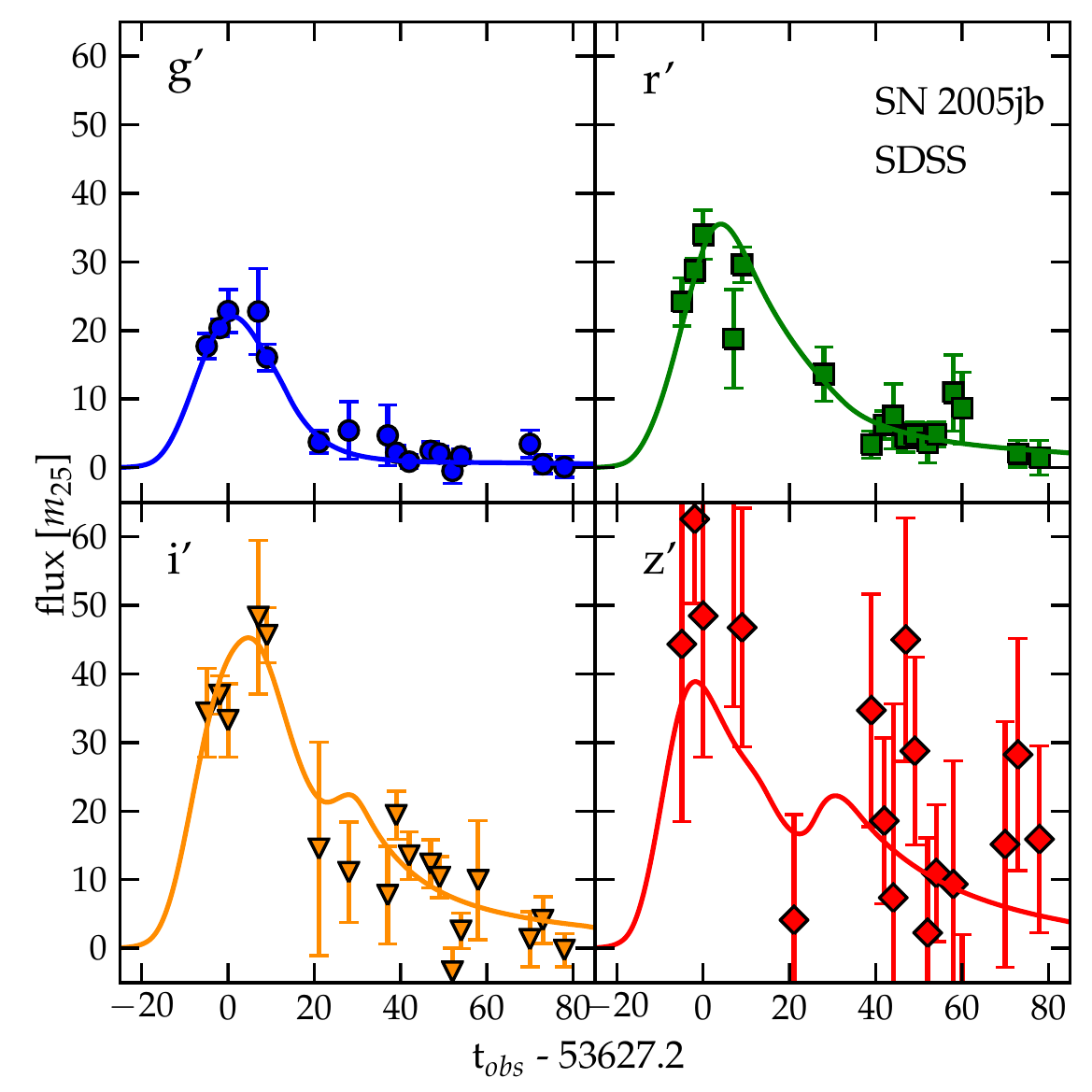}
\else
  \includegraphics[width=0.45\textwidth]{\figdir/f05a_gray}
  \includegraphics[width=0.45\textwidth]{\figdir/f05b_gray}
\fi
\caption{ Two examples of the maximum likelihood SOFT template matches
  for real SN light curves from the verification data. 
  {\em Left:}
  SN~2005jb from the SDSS  survey, with a measured redshift of
  z=0.258.  The best-fitting light curve from the SOFT library is
  based on the normal Type Ia SN~1996X. Note that the $u'$ band
  is not shown.
  {\em Right:}
  SNLS-03D4cz from the SNLS survey. The
  spectroscopic redshift of this object is z=0.695, and the maximum
  likelihood match from SOFT is based on SN~2000cn. 
}
\label{fig:fitExamples}
\end{figure*}

In Figure~\ref{fig:fitExamples} we show two examples of the real SN
light curves used in these verification tests.  The first is SN~2005jb
from the SDSS survey, which has a spectroscopic redshift measurement
of z=0.258.   The second example is SNLS-03D4cz, at a redshift of
0.695, from the SNLS survey.  Over-plotted lines in this figure show
the SOFT template match that produces the maximum likelihood for each
of these light curves. 

\subsection{Results}
\label{sec:Results}

\begin{figure*}[!tb]
  \centering
  \ifrgb{
    \includegraphics[width=\textwidth]{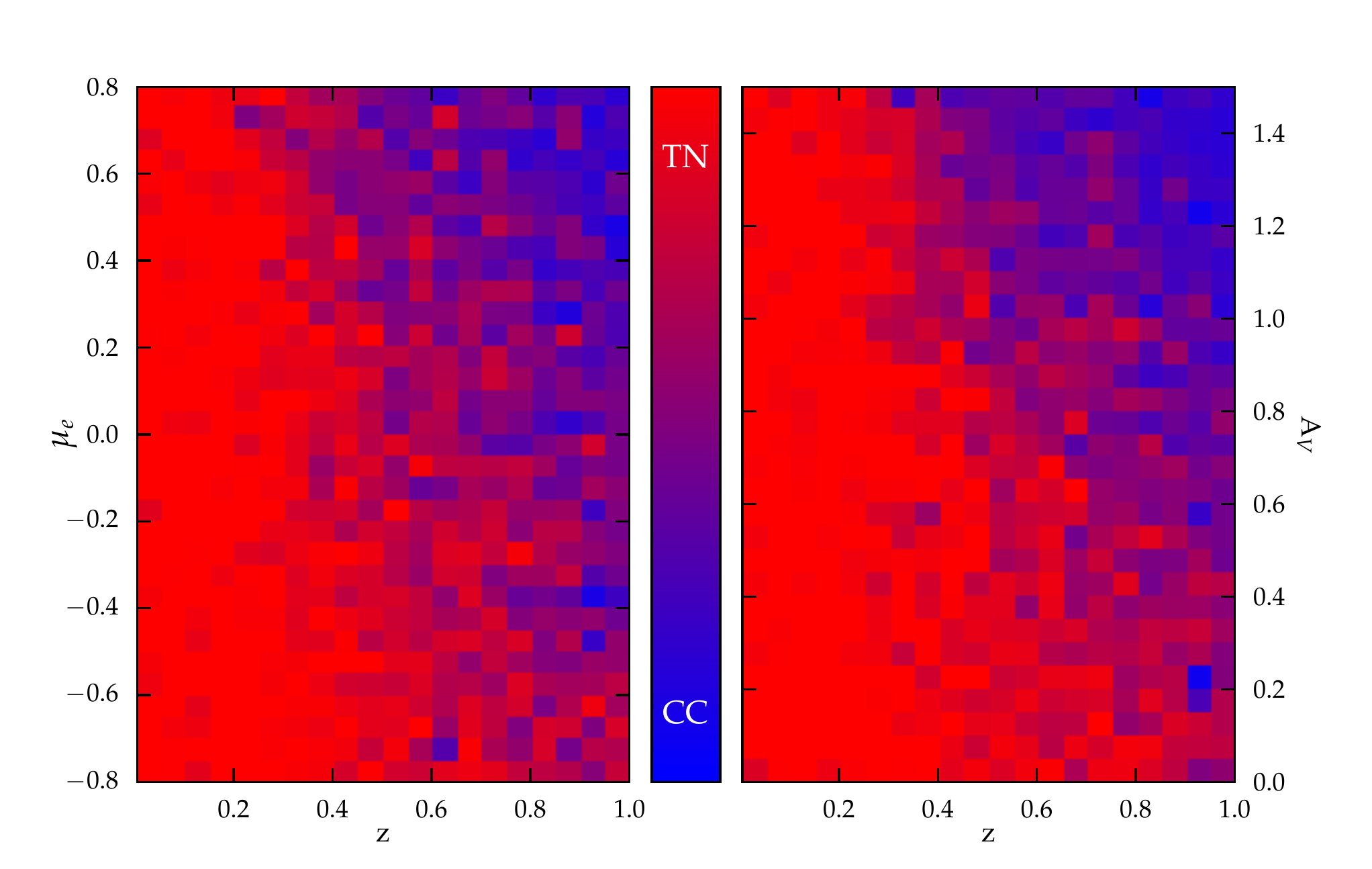} }
  \else{
    \includegraphics[width=\textwidth]{\figdir/f06_gray} }
  \fi
  \vspace{-3mm}
  \caption{ Results from the synthetic light curve classification
    tests.  Using TN SNe in Table~\ref{tab:templib} as models, we
    constructed 5000 synthetic light curves spread uniformly over the
    parameter space.  For each light curve we used the SOFT method to
    calculate $PMG_{TN}$ and $PMG_{CC}$, the normalized composite
    membership grades for the TN SN and CC SN fuzzy sets,
    respectively.  We then divide the parameter space of \mue\ and z
    into bins of size $\Delta$\mue=0.05 mag and $\Delta$z=0.05.  For
    each bin we collect all the synthetic light curves that had input
    values of \mue\ and z drawn from that region of parameter space,
    and then compute the average membership grade returned from that
    bin.  We then set a grayscale value for each bin, with black (red
    in the electronic version) indicating that the average $PMG_{TN}$
    for a bin is 100\%, and white (blue) indicating $PMG_{CC}$=100\%.
    In the right panel we have done the same binning and averaging,
    this time with the y-axis marking \Av.  See the electronic edition
    of the Journal for a color version of this figure.  }
  \label{fig:classTN}
\end{figure*}

\begin{figure*}[!tb]
  \centering
  \ifrgb{
    \includegraphics[width=\textwidth]{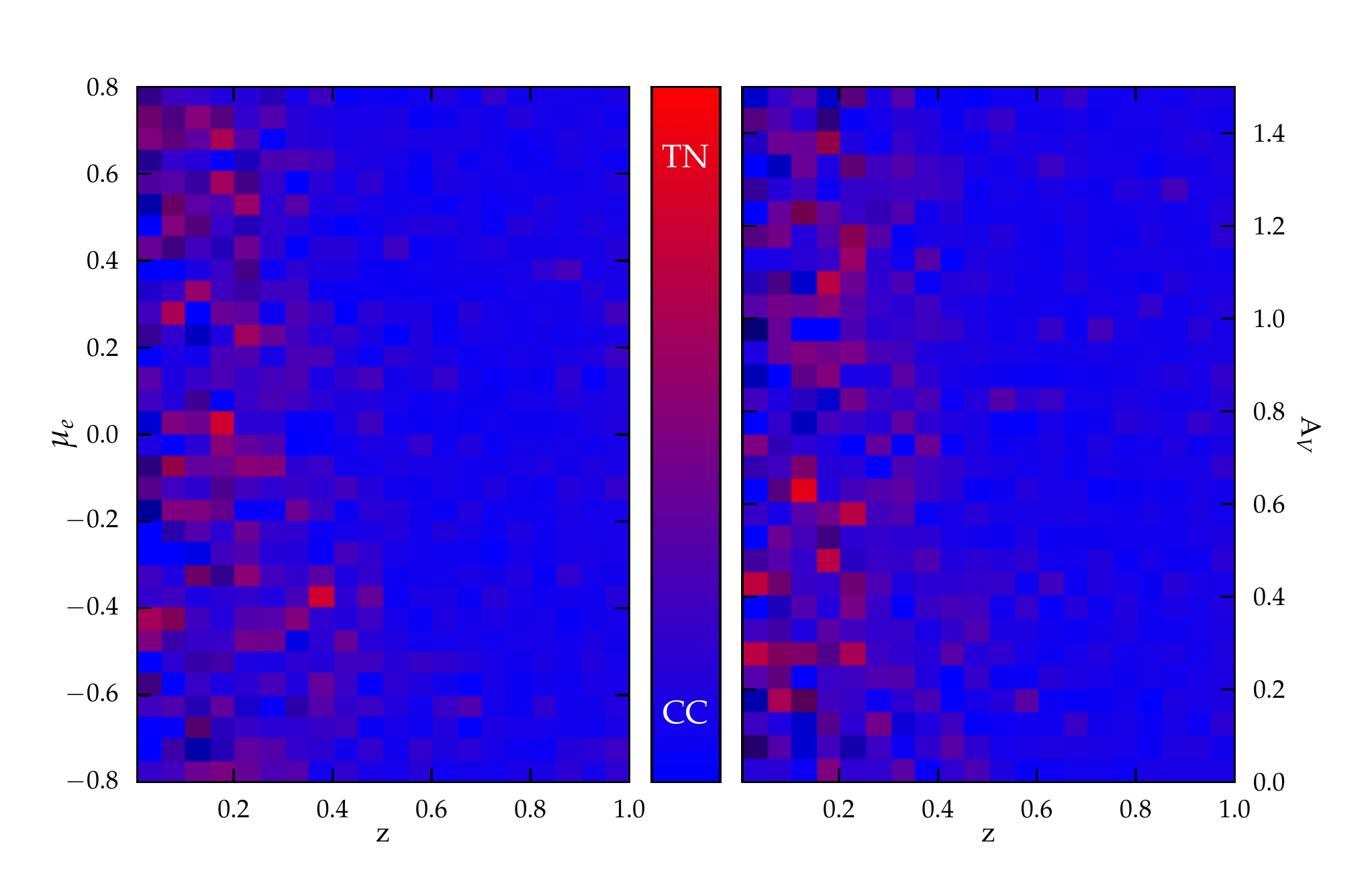}}
  \else{
    \includegraphics[width=\textwidth]{\figdir/f07_gray}}
  \fi
  \vspace{-3mm}
  \caption{ Classification probabilities, as in
    Figure~\ref{fig:classTN}.  In this case we generated 5000 CC SN
    light curves with random parameter vectors using the CC
    templates from Table~\ref{tab:templib} and once again computed the
    normalized fuzzy set classification grades using SOFT.
    As before, black (red in the electronic version) indicates
    $PMG_{TN}$=100\% and white (blue) indicates $PMG_{CC}$=100\%. 
    See the electronic edition of the Journal for a color version of
    this figure. 
  }
  \label{fig:classCC}
\end{figure*}

\begin{figure*}[!tb]
  \centering
  \ifrgb{
    \includegraphics[width=0.49\textwidth]{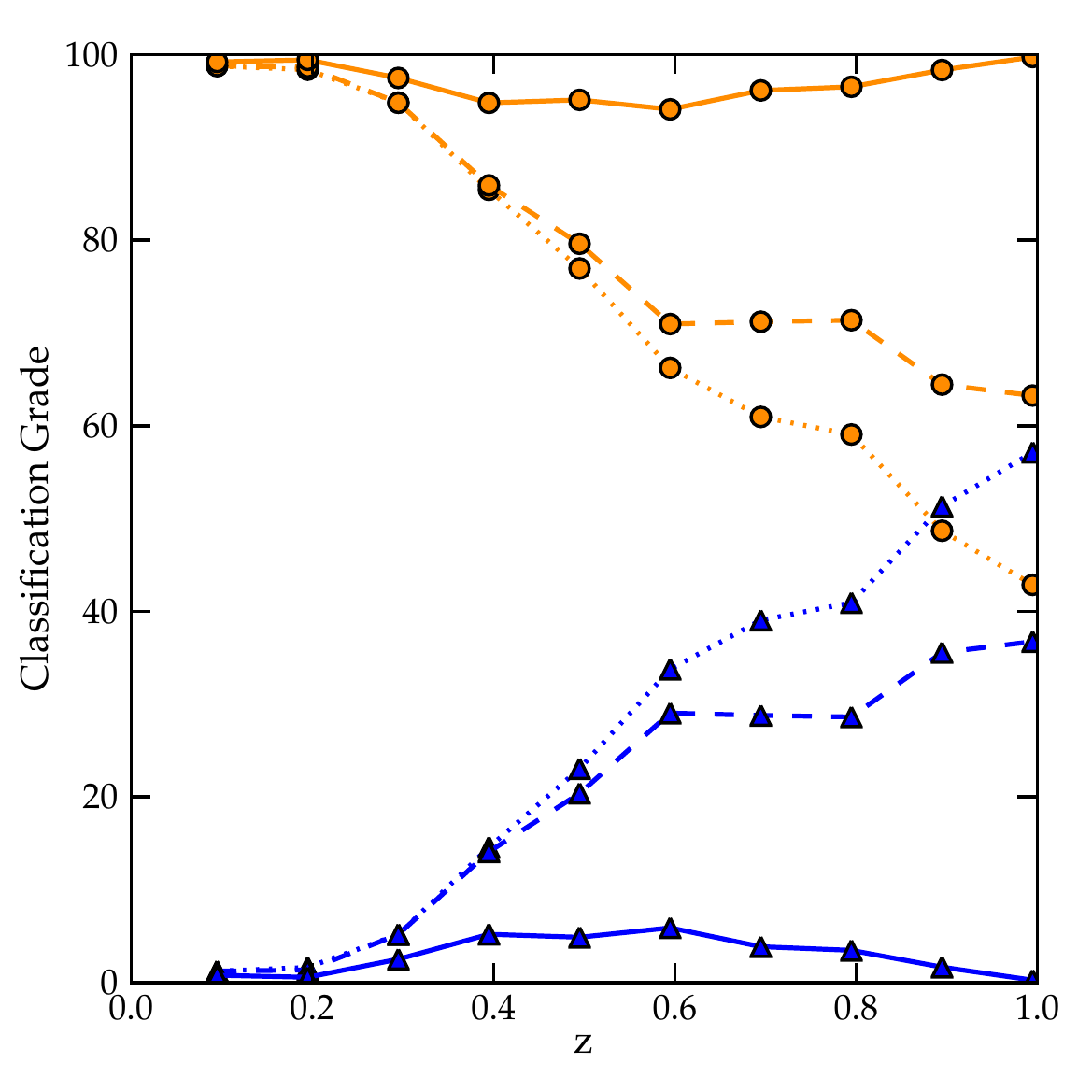}
    \includegraphics[width=0.49\textwidth]{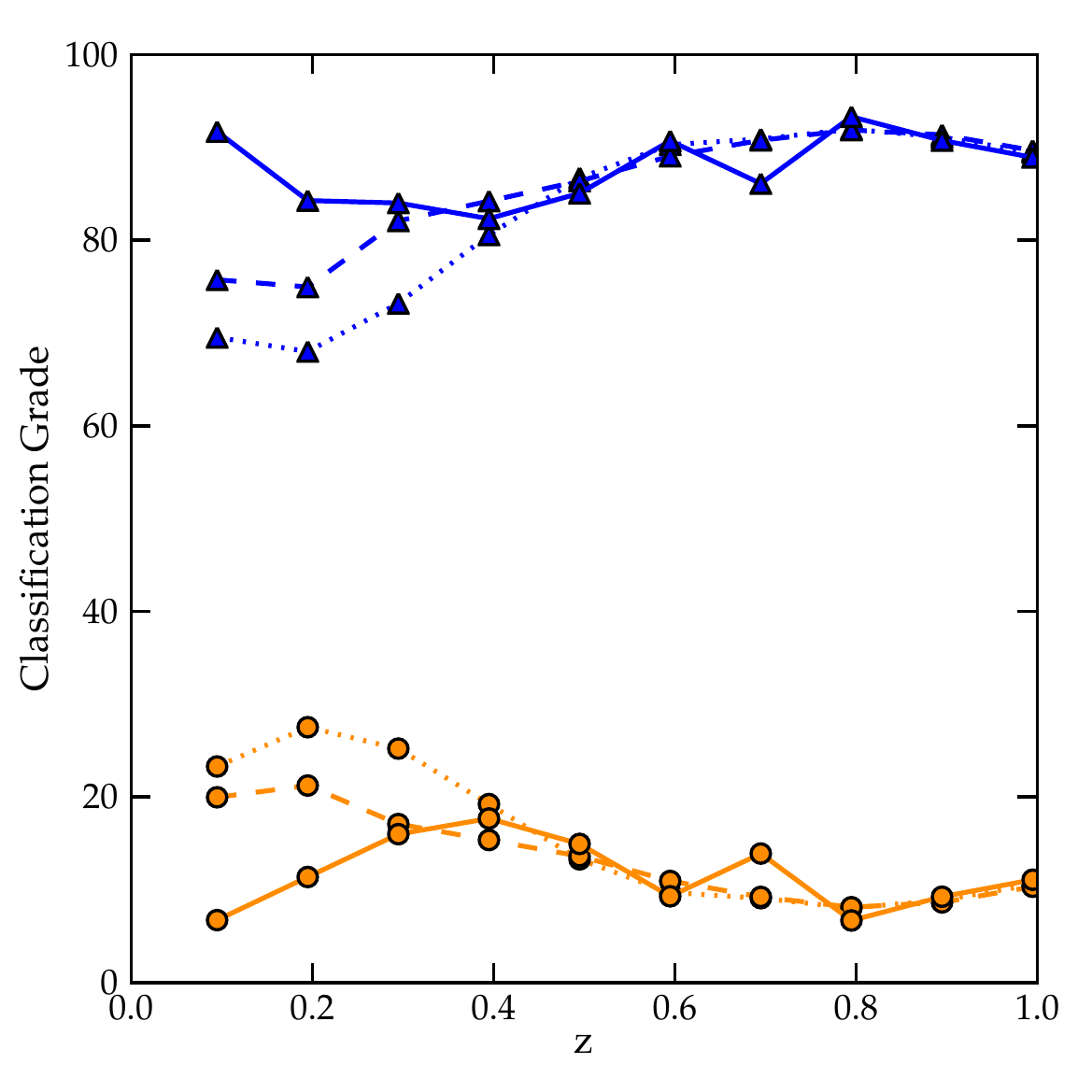}   }
  \else{
    \includegraphics[width=\columnwidth]{\figdir/f08a_gray}
    \includegraphics[width=\columnwidth]{\figdir/f08b_gray}  }
  \fi
  \caption{ Mean classification probabilities as a function of redshift.
    Here we have summed over all \mue\ and \Av\ values, collapsing the
    grids from  Figs.~\ref{fig:classTN} and~\ref{fig:classCC}. 
    The left panel shows classification results for 5000
    synthetic light curves based on TN SN models.  The right panel
    shows the CC SN results.  In both panels the (orange) circles show
    the average $PMG_{TN}$ value as a function of z and the (blue)
    triangles indicate the average $PMG_{CC}$. 
    In each case the dotted lines show mean classification grades from
    all 5000 synthetic SNe.  Dashed lines are from a subset of SNe
    that have a peak signal to noise ratio above 3.  The solid lines
    are constructed from SNe with signal to noise above 10.
    See the electronic edition of the Journal for a color version of
    this figure. 
  }
  \label{fig:classLines}
\end{figure*}

The classification results from our Monte Carlo simulations are shown
in Figures~\ref{fig:classTN} and \ref{fig:classCC} for the \TNSN\ and
\CCSN\ model templates, respectively.  The left-hand panel of each
figure shows the z-\mue\ plane divided into a regular grid.  In each
box on the grid, we collected all the synthetic light curves whose
input parameters (z,\mue) fell within that box. There are 400 boxes,
so each box contains roughly a dozen of the 5000 synthetic light
curves, and each of those dozen light curves has a slightly different
\mue\ and z, and perhaps a very different set of input values \Av\ and
\tpk.  Thus, each contributing synthetic SN light curve yields a
different set of classification grades $PMG_{TN}$ and $PMG_{TN}$.
Combining all contributing light curves for a given box, we can
compute the average classification grade for each of the two principal
classes.  The resulting classification grade image shown in Figures
\ref{fig:classTN} and \ref{fig:classCC} is black in regions where most
synthetic SNe are classified as TN SNe, and turns white in regions
where most are found to be CC SNe.  Another representation of these
data is given in Figure~\ref{fig:classLines}, which shows the average
classification grade as a function of redshift for both the TN and CC
model sets. 

In Figure~\ref{fig:classTN} the input light curves are based on the TN
model subset, so a perfectly accurate classifier would produce a
completely black image, by returning a 100\% probability that every
synthetic light curve is a TN SN. 
The SOFT classifier is not perfect, so most of the boxes in the 
left panel have average $PMG_{TN}$ values that are less than 100\%,
particularly in the upper right corner, for high values of redshift,
distance, and extinction. In Figure~\ref{fig:classCC}, the regions
where the input CC models are imperfectly classified with some
non-zero value for $PMG_{TN}$ appear gray instead of white.  Almost
all of this contamination arises from light curves based on Type Ib/c
models. 

\begin{figure}[!tb]
  \centering
  \ifrgb{ 
    \includegraphics[width=0.48\textwidth]{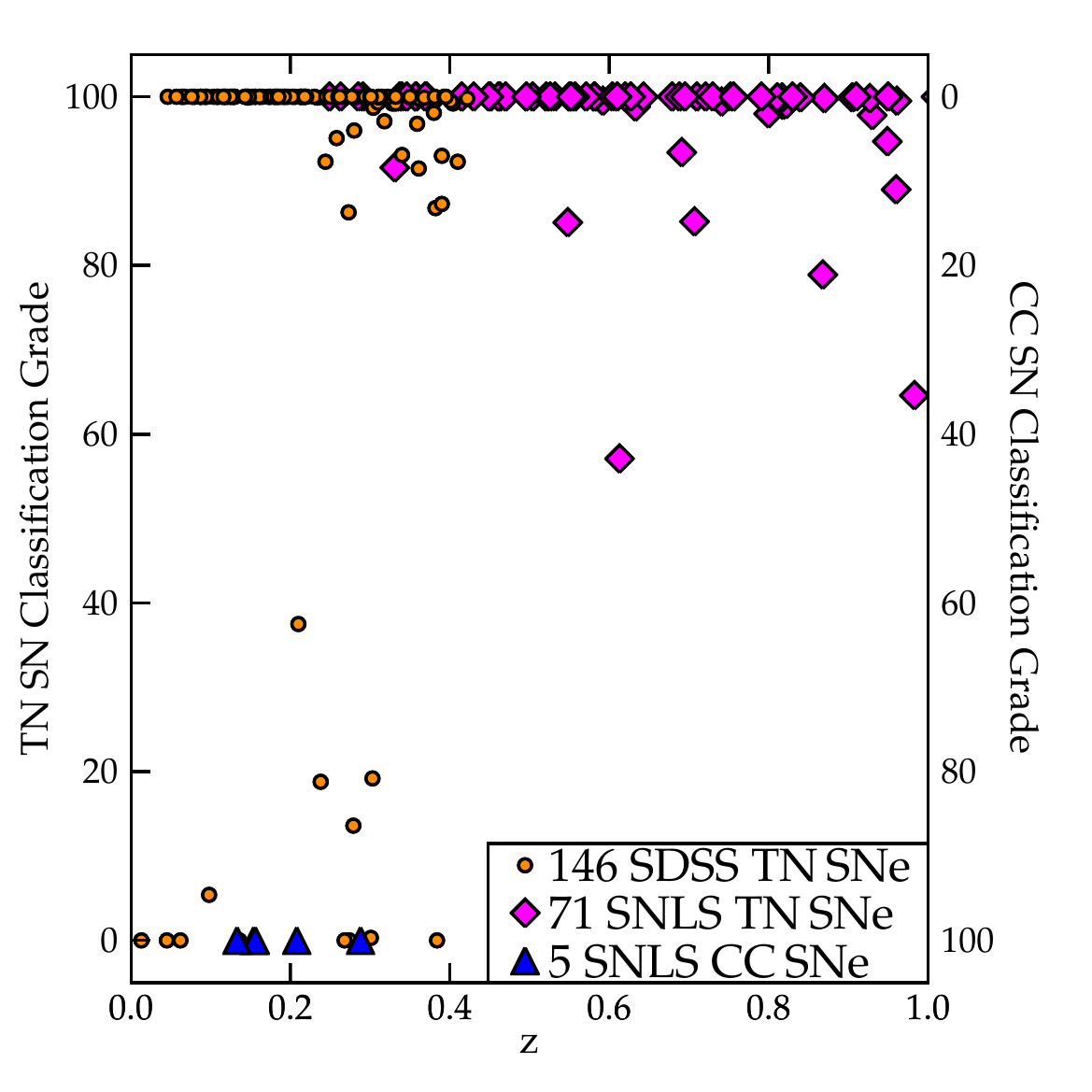} }
  \else{
    \includegraphics[width=0.48\textwidth]{\figdir/f09_gray} }
  \fi
  \caption{ 
    Classification results for 222 SNe from the SDSS and SNLS surveys.  
    The y axis carries the normalized SOFT classification grade, 
    from Equation~\ref{eqn:PMGTN}. The x axis
    marks the published spectroscopic redshift, from either the SN
    itself or its host galaxy.  146 Type Ia SNe from the SDSS survey
    are plotted as (orange) circles, the 71 Type Ia SNe from SNLS are
    shown as (magenta) diamonds, and the (blue) triangles indicate 5
    Type IIP objects from SNLS (note that two of these IIP SNe have
    almost identical redshift and classification grades, so their
    triangles are indistinguishable). 
  }
  \label{fig:classReal}
\end{figure}

Of the 5000 light curves that
contribute to Figure~\ref{fig:classTN}, SOFT calculates a $PMG_{TN}$
classification grade that is greater than the $PMG_{CC}$ grade for 76\%
of them.  
To limit the false positives and false negatives from the
classification results, we can apply a simple cut to remove the
most problematic objects by rejecting light curves with a peak signal
to noise (S/N) ratio that falls below some threshold.  
Figure~\ref{fig:classLines} shows the mean classification grades 
as a function of redshift for all SNe  (dotted lines)  
as well as the classification grades from subsets with S/N$>3$
(dashed) and S/N$>10$ (solid). 
After applying the still very modest S/N$>10$ cut, the fraction of
\TNSNe\ that are correctly identified by SOFT increases to 98\%.  For
this S/N$>10$ subset, the mean \TNSN\ classification grade is $>90\%$
across all redshifts. 

For the synthetic \CCSNe, the S/N$>10$ cut ensures an average
$PMG_{CC}$ classification grade above $80\%$ across all redshifts. 
The primary source for misclassifications of synthetic
\CCSN\ light curves is the Type Ib/c objects.  In particular, the SOFT
classifier is often confused by Type Ib/c SNe at low redshift that
have a very low S/N (i.e. those with heavy host
galaxy extinction or a very large synthetic distance modulus \mue). 

The results from our verification test of the SOFT classifier using
222 real SNe are shown in Figure~\ref{fig:classReal}.  
As was done in the Monte Carlo simulations, we applied the SOFT
classification program to all of these SNe using a uniform
redshift prior (i.e. we ignore any host galaxy or SN spectroscopy), as
well as a uniform prior on \mue\ (i.e. we do not assume any
cosmology). 
For the combined set of 217 Type Ia SNe from SDSS and SNLS, we found
that 204 of these objects (94\%) returned classification grades that
favor a \TNSN\ classification.  The 13 objects that SOFT
misclassified were all from the SDSS survey, and in most cases we 
can deduce the reason for the error.  

Six of the thirteen misclassified objects were given a spectroscopic
classification of ``SNIa?'' by \citet{Holtzman:2008}, indicating some
classification ambiguity even when their spectroscopic information is
considered.  Two more have incomplete light curves, with only one or
two observation epochs beyond the light curve peak.  
Two of the misclassifications are objects with extremely broad light
curves.  SN~2005mo and the SDSS object SN7017 (no IAU designation
available) both exhibit a decline rate substantially slower than any
of the objects in our template library. Using the SALT2 light curve
fitter \citep{Guy:2007} as an independent check, we measured values
for the X1 (stretch) parameter of 3.0 for 2005mo and 5.0 for 7017.
For comparison, these values are more than two times larger than the 
slowest-declining Ia$+$ objects in our template library
(SN~1991T,90N,99gp).  Furthermore, these SALT2 X1 values are larger
than 98\% of the objects in the CfA3 ``constitution set'' of
\citet{Hicken:2009a}.  Unlike light curve fitters that parameterize
the light curve shape (such as SALT2), the SOFT classifier is unable
to extrapolate to light curves with shapes well beyond the limits of
its template library. In order to correctly classify all of these
extremely broad Type Ia SNe, we would need to include some very broad
Type Ia template light curves in the library. 

 A final outlier is the very peculiar SN~2005gj,
which SOFT classifies as a Type IIn object.  This object is an example
of the rare sub-class of hybrid Type Ia/Type IIn objects, exhibiting
spectral indications of a substantial circumstellar medium
\citep{Aldering:2006}.  Although we included templates of some
peculiar Type Ia SNe in the SOFT library (SN~2000cx and 2002cx),
we do not have any representatives of this hybrid ``IIa''
class.

\section{Other Models}\label{sec:OtherModels}

The use of fuzzy set theory in the SOFT method allows us to escape
from the assumption that our set of SN templates describes all
possible light curve shapes. There is, however, an additional
assumption that we have not addressed: the SOFT method still makes the
implicit assumption that all objects being classified are already 
known to be SNe.   In our verification tests this assumption is
perfectly valid, because the Monte Carlo simulation only generates
synthetic SN light curves, and the published light curves have
all been spectroscopically confirmed as real SNe.   
When selecting SN candidates from a real survey this assumption is
often still justified because the set of all detected transients is
passed through a series of preliminary 
filters to remove objects that are not likely to be SNe. 
For example, variable stars can mimic a SN light curve, but these can
be removed by rejecting transients that are coincident with a stellar
source in static sky images, as well as any transients that show
clearly periodic brightness fluctuations. However, even with
aggressive rejection thresholds, some non-SN objects may slip through
to a candidate list.  Furthermore, if potential SN candidates are
vigorously rejected before attempting classification, then some actual
SNe are likely to be unintentionally discarded, too. If this
possibility is not accounted for, then any task which requires a
complete supernova sample (such as a SN rate calculation) will give
misleading results. 

The most thorough way to address this problem in the BATM/SOFT
framework would be to add a new model for every possible class of SN
impostor.  In principle, one could construct 
light curve templates for RR Lyrae, Cataclysmic Variables, Active
Galactic Nuclei, stellar flares, etc.   Each light curve template
could be adapted with the same location parameters \bftheta\ that
were used for our SN templates, and the mechanics of computing
probability distributions would remain the same from there. 
This solution is computationally
complex, and is hindered by the great diversity of possible transient
sources.  We do not pursue this template-based approach here,
but instead we consider how two non-SN models may be introduced to
catch two specific categories of transient contaminants.

\subsection{Zero Flux Model}\label{sec:ZeroFluxModel}

A first category of non-SN interlopers that are often found in SN
candidate sets are objects with a very low signal to noise ratio
across the light curve.  These objects may have light curves that are
intrinsically very different from SN light curves, but they have
managed to slip through the pre-classification filters because the
data quality is poor enough to allow some ambiguity.  When we
encounter a low-flux light curve, if we assume that it is a SN in
order to apply a SOFT classification, then we are artificially
inflating the SN classification probabilities by ignoring the
possibility that it is some other type of object. 

To address these contaminants within the BATM/SOFT framework, we can
introduce a ``Zero Flux Model'', (ZFM) which asserts that there is no
SN observed, so the true flux of the light curve is zero everywhere.  
The Bayesian likelihood of a candidate match using this model is
similar to Eq.~\ref{eqn:p(D|theta,Mj)}, but without any location
parameters \bftheta, and without the model flux term $\mathcal{F}$:
 
\begin{equation}
   p(\bfD|ZFM) = \prod_{i=1}^N{\frac{1}{\sqrt{2\pi}\sigma_i}
\mbox{exp} \left(\frac{-f_i}{ 2~\sigma_i^2}\right) }
\label{eqn:p(D|ZFM)}
\end{equation}

\noindent To include this likelihood in the set of PMPs for
Eq.~\ref{eqn:PMPj}, we must assign a prior $p(\bfD|ZFM)$.
Ideally, this prior should represent an initial guess at the fraction
of low signal to noise candidates that are false positives. 
A carefully chosen ZFM prior would reflect the details of the data
reduction and candidate selection pipelines.  In practice, it will
generally be sufficient to apply a rough prior that is of the same
order as the SN model priors, letting the data drive the ZFM
probability. 

\begin{figure}[!tb]
\centering
\ifrgb
  \includegraphics[width=\columnwidth]{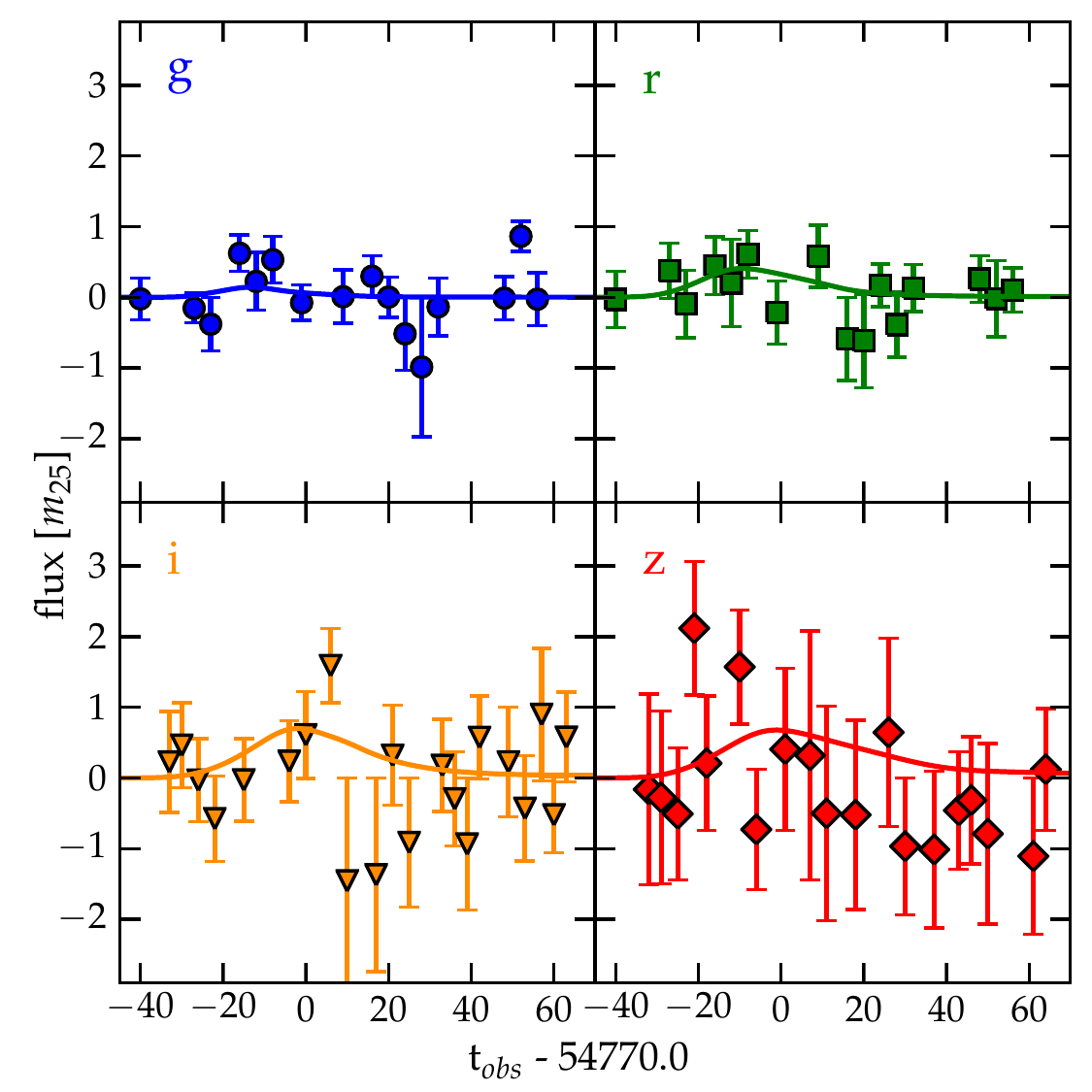}
\else
  \includegraphics[width=\columnwidth]{\figdir/f10_gray}
\fi
\caption{ 
  A synthetic light curve example demonstrating the importance of
  including a zero flux model (ZFM) to weed out over-confidence.  
  This ``light curve'' was generated as pure random noise, and was
  then passed through the SOFT program to measure normalized
  membership grades for classification.  
  If SOFT only considers the SN templates then the classification
  probabilities would be $p(CC|\bfD)=0.91$ and 
  $p(TN|\bfD)=0.09$.  The maximum likelihood SN template match in
  this case comes from the Type Ia SN~1996X, shown here as the solid
  curves.  When the ZFM is included, the renormalized classification
  grades become $p(CC|\bfD)=0.67$, $p(TN|\bfD)=0.06$, and 
  $p(ZFM|\bfD)=0.27$.  
  See the electronic edition of the Journal for a color version of
  this figure. 
 }
\label{fig:zfexample}
\end{figure}

The ZFM is also useful in providing a measure of the confidence that
should be given to SN classifications.
The observed light curve of a very faint SN can often be matched
equally well by many light curve templates because the observational
errors are so large.  BATM and SOFT would then classify the object
based primarily on the model priors defined in
\S\ref{sec:ModelPriors}.  In that case our classification grades
should reflect the fact that the data are providing little or no new
information.  The ZFM model achieves this, by inflating the
normalization factor of Eq.~\ref{eqn:PMPj} whenever the light curve
signal gets buried by observational noise.  

This effect of the ZFM on classification grades is illustrated
in Figure~\ref{fig:zfexample}, where we present a synthetic light curve
that is entirely made up of random noise.  There is no real signal for
SOFT to fit to, but if the only models available to it are SN
templates, then 
SOFT will necessarily classify the ``object'' into one of the
available SN classes.  When the ZFM is included, however, we find a
significant likelihood that there  is no object there at all, and the
renormalized classification grades reflect the ambiguity of this very
low-signal event.  

\subsection{Something Else Model}\label{sec:SomethingElseModel}

The recent discovery of the peculiar optical transient SCP~06F6 with the
Hubble Space Telescope \citep{Barbary:2009} provides a prime example
of another category of non-SN contaminants.  This object has a bright,
clear light curve that is superficially rather similar to a SN, but it
is unlike any known supernova both photometrically and
spectroscopically \citep{Gansicke:2009}.
If a transient similar to SCP~06F6 were detected by
a survey like Pan-STARRS or LSST, its light curve would not be initially
reviewed by human eyes (with thousands of new candidates discovered
per night, there just aren't enough grad students available). After
the SCP~06F6-like candidate passes through preliminary sifting it
would probably be labeled as ``SN-like'' and get passed off to an
automated classification algorithm such as BATM or SOFT.    
If the classifier only compares this object against a set of
SN templates, it will necessarily misclassify the object, no matter
how broad or deep the template library is.   

To account for the possibility that a candidate light curve could be
something that resembles a SN but is not one of the known SN types, we
need a flexible model that can fit any conceivable light curve
reasonably well.  For this purpose, we introduce the 
``Something Else Model'' (SEM), a simple non-physical model designed to
replicate a non-periodic light curve.
The SEM  consists of a set of spline curves that are 
required to pass through a set of 5 magnitude points positioned at
regular intervals in time.  The outer two points are positioned at the
end-points of the observed light curve, and the other three are spaced
evenly over the interior.
The magnitude of each knot $m_k$ is a free parameter that is
allowed to vary over a range $\Delta m_k$ from zero to 30 magnitudes,
with a uniform prior across that region.
This span comfortably encompasses the range of magnitudes accessible
to any SN survey. Since this model is not physically
motivated, we model each bandpass completely independently.  The
posterior likelihood for a given vector of spline knots ${\bf m_K}$ is
calculated in the same manner as Eq.~\ref{eqn:p(D|theta,Mj)}:

\begin{equation}
  \begin{array}{ll}
    p(\bfD|{\bf m_k},SEM) = & 
    \prod_{i=1}^N\frac{1}{\sqrt{2\pi}\sigma_i} \\
      & \times~ {\mbox{exp} \left(\frac{-(f_i -
        \mathcal{F}_{SEM}(t_i,{\bf m_k}))^2}{ 2~\sigma_i^2}\right)}
  \end{array}
  \label{eqn:p(D|mk,SEM)}
\end{equation}

To determine the net likelihood for the SEM, we must integrate this
over all possible spline knot vectors ${\bf m_k}$.

\begin{equation}
    p(\bfD|SEM) = \int\limits_{\bf m_k} p(m_k) p(\bfD|m_k,SEM)
    {\bf dm_k}
  \label{eqn:p(D|SEM)}
\end{equation}

\begin{figure}[!tb]
\centering
\includegraphics[width=\columnwidth]{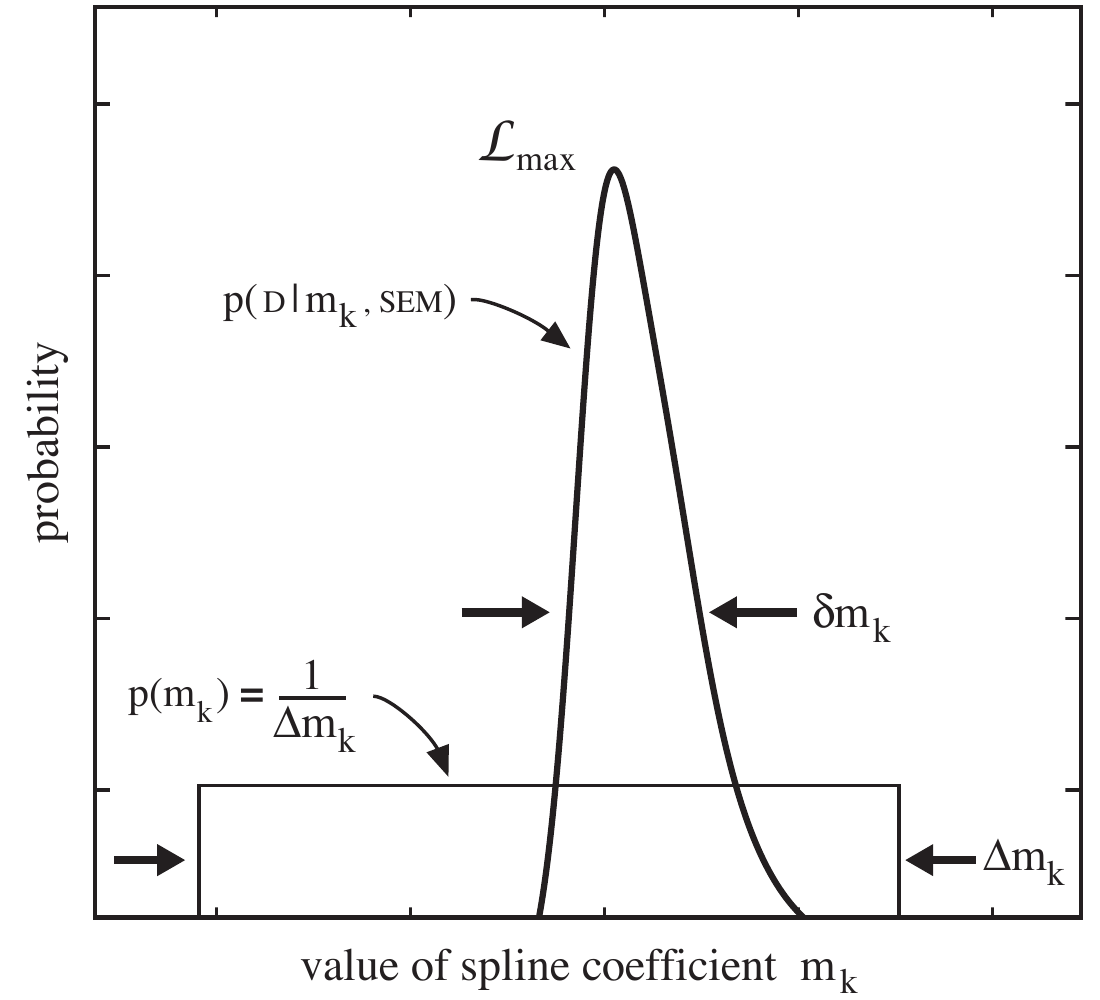}
\caption{ 
  Estimating the area of the SEM likelihood distribution along one
  dimension of parameter space.  
  The curve represents the posterior probability $p(\bfD|m_k,SEM)$.
  The rectangle shows the uniform prior $p(m_k)=1/\Delta m_k$ 
  spanning a wide range of possible spline knot values.  
  A downhill simplex
  minimization determines the value of the spline parameter $m_k$
  that yields the maximum likelihood value $\mathcal{L}_{max}$.  
  Through numerical
  sampling of the likelihood distribution around that point, we can
  estimate the width of the likelihood peak, $\delta m_k$, and get a
  rough approximation for the area under the curve as
  $\mathcal{L}_{max} ~ \delta m_k$. 
  The prior times the posterior probability is then estimated as the
  product of $\mathcal{L}_{max}$ with the Occam 
  factor, $\delta m_k / \Delta m_k$.   This figure adapted from
  \protect\citet{Gregory:2005}. 
 }
\label{fig:occam_dmk}
\end{figure}

To simplify the computation of this integral, we can use a rough
maximum likelihood approximation. By design, there will always be some
set of knots for which this model can provide a reasonably good fit to
any light curve, so the likelihood distribution over the knot space
will be sharply peaked. In such a regime, we can estimate the integral
of Eq.~\ref{eqn:p(D|SEM)} as the area of that single peak.  Let us
denote the height of the peak as the maximum likelihood 
$\mathcal{L}_{max}= \max( p(\bfD|m_k,SEM) )$, and the width of the
peak in the plane of parameter $m_k$ as $\delta m_k$.  A first-order
approximation for the peak area in that plane is $\mathcal{L}_{max}
\delta m_k$.  
To apply a uniform prior over the range $\Delta m_k$ we multiply by a
factor $1/\Delta m_k$, and taking the product over all knots in all
bandpasses, we get 
$\mathcal{L}_{max}  \prod \frac{\delta m_k}{\Delta m_k}$.  This
estimate can then be included as the numerator of
Equation~\ref{eqn:PMPj} to compute the SEM posterior model
probability, $p(SEM|\bfD)$.  The components of
this approximation are illustrated in Figure~\ref{fig:occam_dmk}.
The product $\prod \frac{\delta m_k}{\Delta m_k}$, sometimes referred
to as the ``Occam factor,''  decreases rapidly
as the number of knots increases, and therefore it effectively
punishes the SEM model for using many more parameters than the SN
template models
(for a derivation of this maximum likelihood approximation and further
discussion of the Occam factor, see section 3.5 of
\citet{Gregory:2005}).  

To get $\mathcal{L}_{max}$ we find the maximum likelihood spline fit
for each bandpass using a downhill simplex 
minimization and defining each likelihood value with Equation
\ref{eqn:p(D|Mj)}. We then estimate $\delta m_k$ by sampling the
likelihood distribution in the vicinity of the peak to find the full
width at half maximum (see Fig.~\ref{fig:occam_dmk}). 

The final step to include this SEM in our total probability
calculation is to define an appropriate model prior.  As with the ZFM,
the $p(SEM)$ prior should reflect the data processing pipeline by
providing a low probability when the upstream filtering of
candidates is more stringent.  In general, the occurrence of true
``Something Else'' objects is expected to be very rare, so a value of
$p(SEM)$ that is three or more orders of magnitude less than the SN
model priors would be appropriate.

\begin{figure}[!tbp]
\ifrgb
  \includegraphics[width=\columnwidth]{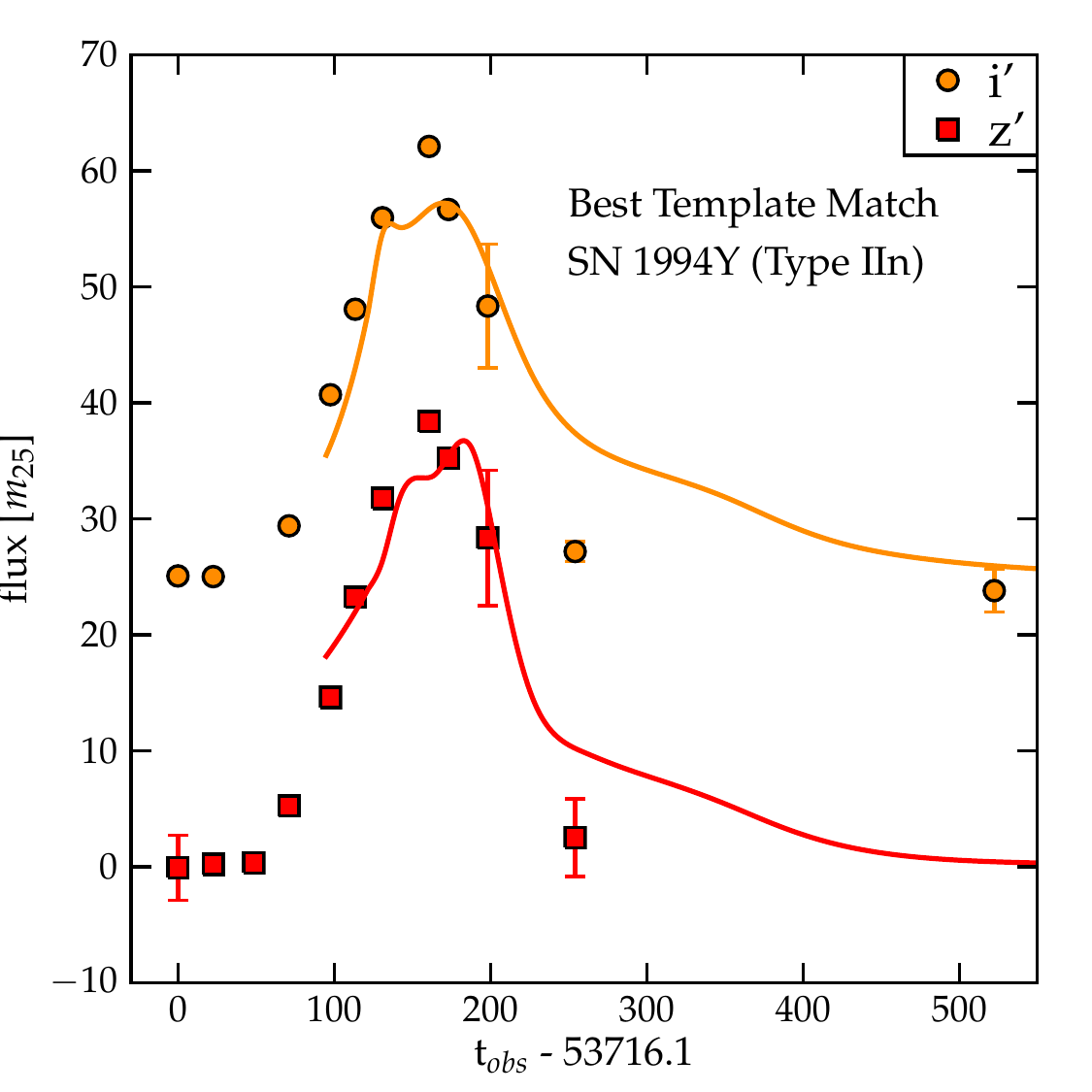}
  \includegraphics[width=\columnwidth]{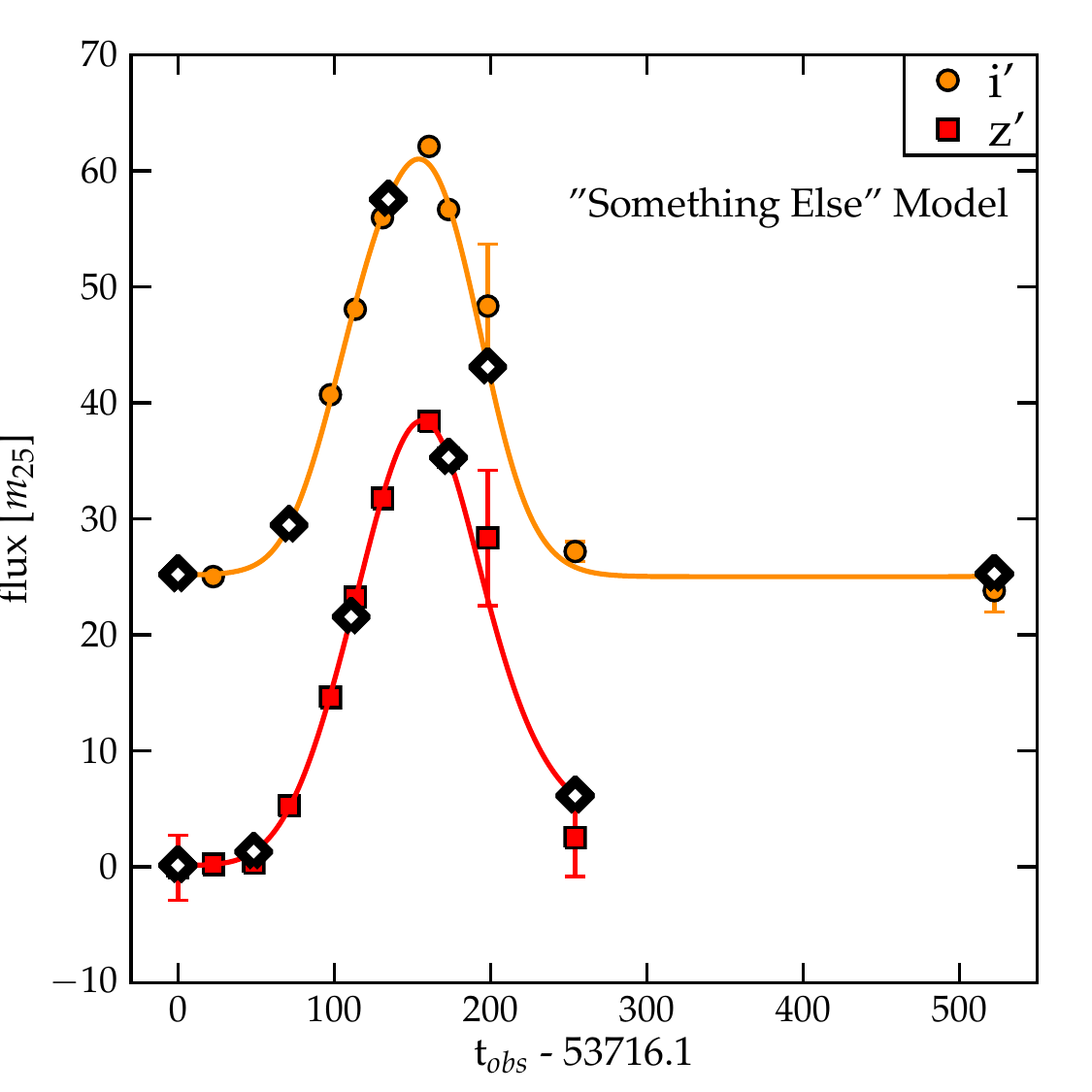}
\else
  \includegraphics[width=\columnwidth]{\figdir/f12a_gray}
  \includegraphics[width=\columnwidth]{\figdir/f12b_gray}
\fi
\caption{ Light curve fits for the unusual transient object SCP~06F6.
  The data points from \protect\citet{Barbary:2009} are shown as
  circles for the the HST F775W ($i'$) band, and as squares for F850LP
  ($z'$).  {\em Top:} The best fitting template is the Type IIn
  SN~1994Y, at a redshift of 0.2, but even this model provides a very
  poor fit.  {\em Bottom:} The SEM is capable of providing a very good
  fit to these data.  The diamond symbols mark the locations of the
  spline knots that define the SEM fit.  In this case the likelihood
  of a match for the SEM is larger than that from any of the SN
  templates.  }
\label{fig:somethingelse}
\end{figure}

In Figure \ref{fig:somethingelse} we demonstrate the application of
this SEM using the two-color SCP 06F6 light curve of
\citet{Barbary:2009}. If we try to determine the probability that this 
object is a CC or TN SN, then BATM and SOFT measure low values for
both $p(CC)$ and $p(TN)$. 
If we leave out the SEM comparison, then when these low likelihood
values are normalized against themselves using Eq.~\ref{eqn:PMPj},
the  result is a pair of grossly exaggerated probabilities.   By
including the SEM case, BATM and SOFT can report with high confidence
that this object is unlike both the CC and the TN classes. 

The structure of the SEM outlined here is necessarily very loose. Due
to the relatively large number of parameters and the wide open
parameter space, this model will usually reach a reasonably high 
maximum likelihood $\mathcal{L}_{max}$.  Those same flexibility
features, however, will then substantially reduce the posterior
probability through the Occam factor, because the characteristic width
$\delta m_k$ is generally much smaller than the parameter range
$\Delta m_k$. 
The size of the SEM prior, the allowed parameter range $\Delta m_k$,
the number of spline knots per filter 
and their spacing in time are all important components in determining
the posterior probability $p(SEM|\bfD)$.  These values should be
tuned for any particular survey by fitting the SEM to known SN
transients as well as examples of possible non-SN contaminants.  A
suitably optimized SEM will yield a relatively low $p(SEM|\bfD)$
for real SNe, while still providing a non-zero $p(SEM|\bfD)$ for
unusual interlopers.

\section{Conclusions and Future Work}
\label{sec:Conclusions}

We have presented in detail the updated BATM (Bayesian Adaptive
Template Matching) program, and showed that a strictly probabilistic
framework is insufficient for SN classification using fixed-shape
templates.  To overcome the practical and theoretical shortcomings of
the BATM approach we have introduced SOFT, a method based on Fuzzy Set
Theory.  This technique allows us to classify SNe into the broad
categories of thermonuclear (TN) and core-collapse (CC) SNe using
photometry alone.  

The SOFT method is demonstrated on simulated observations of synthetic
SNe modeled after the Pan-STARRS 1 (PS1) survey.  Sampling a wide
range of redshift, distance and extinction we find that SOFT can
produce average classification accuracies above 90\% for \TNSNe\ and
above 80\% for \CCSNe.  Additional verification of the SOFT
classification accuracy is presented with 146 \TNSN\ light curves from
the SDSS-II survey, which again yield better than 90\% classification
accuracy.  Almost all of the SDSS SN misclassifications could be
influenced by peculiarities and ambiguities in the light curves
themselves. 

We suggest two additional non-SN models that can be helpful in
identifying anomalous objects within a SN candidate set.  The Zero
Flux Model catches SN impostors that have a very low signal to noise
ratio, and the Something Else Model provides a flexible, non-physical
model to assist in the identification of unusual objects like the
peculiar transient SCP 06F6. 

A number of further refinements to the SOFT program could improve its
functionality and accuracy for automated SN classification.  The
strength of any template matching software is 
limited by the size and quality of the template library it draws from.
The set of templates used for this work has been much enhanced in this
regard since the earliest incarnation of BATM, but there is much room
for further development.  In particular, collecting template light
curves with more complete UBVRI data -- especially in the rise time of
the light curve --  would allow for more accurate spline fits to
define the primary templates. This is particularly important for the
CC SNe, for which the available collection of well-sampled light
curves and early-time spectra is notably thin.  Adding to the template
library some light curves observed at high redshift could provide a
useful supplement to the scarce UV data from the low-z sample.

The SOFT method introduces two new parameters that characterize the
template library itself: $\sigma'_j$ quantifies the amount of
``fuzziness'' and the $q$ parameter controls the strength of the Fuzzy
combination operator.  These parameters reflect the number and
diversity of templates in the template library, and appropriate values
can be estimated from a comparison of templates against each other.  We
performed a preliminary calibration of these parameters and settled on
$\sigma_{TN}=0.1$ for \TNSN\ templates, $\sigma_{CC}=0.15$ for
\CCSN\ templates, and q=0.2 for all combinations.   A more precise
determination of $\sigma'_j$ and q using a large training set of
well-observed SNe would be a useful next step.

In addition to classifying SNe, the SOFT method can be used
to determine parameter estimates for candidate SNe, based on the
comparison of light curves. This extension of the SOFT technique is
presented in a companion paper (Rodney and Tonry, 2009), where we
focus on the use of SOFT to derive redshift and distance estimates
from \TNSNe\ for constraining cosmological models.

{\bf Acknowledgments:}
We thank the anonymous referee for constructive comments that led to 
significant improvements in this work. 
We thank Stephane Blondin for providing access and assistance with the
collection of SNID spectra and Saurabh Jha for sharing low-z light
curves.  We thank Brian Connolly for helpful discussion and
review of an early draft, and Kathleen Robertson for assistance with
reference materials.   This work made use of the SUSPECT database of
SN light curves and spectra, maintained by Jerod Parrent
 (\url{http://bruford.nhn.ou.edu/~suspect}).

\bibliographystyle{apj}
\bibliography{bibdesk}


%
%




\end{document}
